\documentclass[mathleft, fleqn, final
]{an}
\usepackage{graphicx}
\usepackage{times}
\usepackage{amsmath}
\usepackage[authoryear]{natbib}
\bibpunct{(}{)}{;}{a}{}{,}
\begin{document}

\Pagespan{1}{16}
\Yearpublication{2015}%
\Yearsubmission{2015}%
\Month{-}%
\Volume{-}%
\Issue{-}%

\title{The Gaia spectrophotometric standard stars survey --- II. \\
Instrumental effects  of six ground-based observing campaigns.\thanks{Based on data obtained with:
BFOSC@Cassini in Loiano, Italy; EFOSC2@NTT in La Silla, Chile; DOLORES@TNG in La
Palma, Spain; CAFOS@2.2m in Calar Alto, Spain; LaRuca@1.5m
in San Pedro M\'artir, Mexico (see acknowledgements for more details).}}

\author{G.~Altavilla\inst{1}\fnmsep\thanks{Corresponding author:
  \email{giuseppe.altavilla@oabo.inaf.it}\newline}
\and   S.~Marinoni\inst{2 \and 3}
\and   E.~Pancino\inst{1 \and 3}
\and   S.~Galleti\inst{1}  
\and   S.~Ragaini\inst{1}  
\and   M.~Bellazzini\inst{1}
\and   G.~Cocozza\inst{1} 
\and   A.~Bragaglia\inst{1}
\and   J.~M.~Carrasco\inst{4}
\and   A.~Castro\inst{5}  
\and   L.~Di~Fabrizio\inst{6}
\and   L.~Federici\inst{1} 
\and   F.~Figueras\inst{4} 
\and   M.~Gebran\inst{7}   
\and   C.~Jordi\inst{4}    
\and   E.~Masana\inst{4}   
\and   W.~Schuster\inst{5} 
\and   G.~Valentini\inst{8}
\and   H.~Voss\inst{4}
}
\titlerunning{ Instrumental effects of six ground-based observing campaigns }
\authorrunning{G.~Altavilla et al.}
\institute{
INAF-Osservatorio Astronomico di Bologna, Via Ranzani 1, I-40127 Bologna, Italy
\and 
INAF-Osservatorio Astronomico di Roma, Via Frascati 33, I-00040, Monte Porzio 
Catone (Roma), Italy
\and 
ASI Science Data Center, via del Politecnico SNC, I-00133 Roma, Italy\
\and 
Departament d'Astronomia i Meteorologia, Institut de Ci\`ences del Cosmos
(ICC), Universitat de Barcelona (IEEC-UB), Mart\'\i\  i Franqu\`es, 1, 08028 
Barcelona, Spain
\and 
Observatorio Astron\'omico Nacional, Universidad Nacional Aut\'onoma de
M\'exico, Apartado Postal 106, C. P. 22800 Ensenada, BC, Mexico
\and 
Fundaci\'on Galileo Galilei -- INAF, Rambla Jos\'e Ana Fern\'andez P\'erez,
7, 38712 Bre\~na Baja, Tenerife, Spain
\and 
Department of Physics and Astronomy, Notre Dame University Louaize, PO Box
72, Zouk Mikael, Zouk Mosbeh, Lebanon
\and 
INAF-Osservatorio Astronomico di Teramo, Via Mentore Maggini snc, Loc.
Collurania, 64100 Teramo, Italy}
\received{2015}
\accepted{2015}
\publonline{later}


\abstract{
The Gaia SpectroPhotometric Standard Stars (SPSS) survey started in 2006, it was
awarded almost 450 observing nights, and accumulated almost 100\,000 raw data
frames, with both photometric and spectroscopic observations. Such large
observational effort requires careful, homogeneous, and automated data reduction
and quality control procedures. In this paper, we quantitatively evaluate
instrumental effects that might have a significant (i.e., $\geq$1\%) impact on
the Gaia SPSS flux calibration. The measurements involve six different
instruments, monitored over the eight years of observations dedicated to the
Gaia flux standards campaigns: DOLORES@TNG in La Palma, EFOSC2@NTT and ROSS@REM
in La Silla, CAFOS@2.2m in Calar Alto, BFOSC@Cassini in Loiano, and LaRuca@1.5m
in San Pedro M\'artir. We examine and quantitatively evaluate the following
effects: CCD linearity and shutter times, calibration frames stability, lamp
flexures, second order contamination, light polarization, and fringing. We
present methods to correct for the relevant effects, which can be applied to a
wide range of observational projects at similar instruments.   
}

\maketitle

\section{Introduction}

Gaia\footnote{http://www.cosmos.esa.int/web/gaia/home} is a cornerstone mission
of the ESA (European Space Agency) Space Program, launched in 2013 December 19. 
The Gaia satellite is performing an all-sky survey to obtain pa\-rallaxes and proper
motions to $\mu$as precision for about 10$^9$ objects down to a limiting
magnitude of $V\simeq 20$~mag, and astrophysical parameters ($T_{\mathrm{eff}}$,
$\log g$, $E(B-V)$, metallicity, etc.) plus 2-30 km~s$^{-1}$ precision ---
depending on spectral type --- radial velocities for several millions of stars
with $V<16$~mag. Such an observational effort will have an impact on all
branches of astronomy and astrophysics, from solar system objects to distant
QSOs (Quasi Stellar Objects, or Quasars), from the Galaxy to fundamental
physics. The exquisite quality of Gaia data will allow a detailed reconstruction
of the 6D spatial structure and velocity field of the Milky Way galaxy within
$\simeq 10$ kpc from the Sun, providing answers to long-standing questions about
the origin and evolution of our Galaxy, from a quantitative census of its
stellar populations, to a detailed characterization of its substructures, to the
distribution of dark matter. Gaia will also determine direct geometric distances
to many kinds of distance standard candles, setting the cosmological distance
scale on extremely firm bases. The Gaia scanning law will cover the whole sky
repeatedly \citep[$\simeq$70 times on a\-ve\-rage, over the planned 5-years Gaia
lifetime, see][]{carrasco07,lindegren08}, therefore, a large number of transient
alerts will be a natural byproduct \citep[for example, many SNe will be
discovered, see][]{altavilla12}\footnote{Indeed, Gaia started already
discovering SNe, as testified by the ATel (http://www.astronomerstelegram.org/)
telegrams published so far.}. For a review of Gaia science cases, see
\citet{perryman01}, \citet{mignard05}, and \citet{prusti11}. The challenges in
the data processing of this huge data volume require a team of hundreds of
scientists and engineers \citep[organized in the DPAC, or Data Analysis and
Processing Consortium, see][]{mignard08}.

Gaia is usually described as a self-calibrating mission, but it also needs
external data to fix the zero-point of the magnitude/flux system
\citep{pancino12} and radial velocities \citep{soubiran13}, and to train the
classification and parametrization algorithms \citep{bailerjones13}. Gaia
photometry will come from the astro\-me\-tric array of CCDs in the Gaia G-band,
a filter-less band whose profile is defined by the reflectivity of the mirrors
and by the sensitivity of the detectors, and from the blue (BP) and red (RP)
spectro-photometers which will produce dispersed i\-ma\-ges with
20$<\lambda/\Delta\lambda<$100 over the spectral ranges 330-680 nm and 640-1050
nm, respectively \citep[see also][]{jordi10,prusti12}. The derived
spectro-photometry is essential not only to properly classify stars or estimate
interstellar extinction, but also to compute the chromaticity correction to
centroids of stars, a fundamental piece of information to achieve the planned
astrometric performance \citep{busonero06,lindegren08}. 

The final conversion of internally-calibrated G instrumental magnitudes and
BP/RP instrumental fluxes into physical units requires an external absolute flux
scale, that our team is in charge of providing \citep{pancino12}. The ideal Gaia
spectrophotometric standard stars (SPSS) grid should comprise of the order of
100 SPSS in the range $9\le V\le15$~mag, properly distributed in the sky to be
observed by Gaia as many times as possible. The mission requirement is to
calibrate Gaia data with an {\em accuracy} of a few percent (1--3\%) with
respect to Vega \citep{bohlin07}. We use as {\em Pillars} the three pure
Hydrogen WDs (White Dwarfs) adopted by \citet{bohlin95} as fun\-da\-men\-tal
calibrators. Because the {\em Pillars} are not always visible, we are also
building an inter\-me\-diate grid of calibrators ({\em Primary SPSS}) that are
observable all year round, with 2--4~m class telescopes. 

A set of more than 100 flux tables with $\le$1\% internal consistency and
$\le$1--3\% absolute calibration with respect to Vega are of interest for many
scientific applications \citep[see][for a review of spectrophotometric methods
and catalogues]{bohlin14}, including dark energy surveys based on type Ia
supernovae \citep{sullivan11}. Our observing campaign to build a grid of SPSS,
for the flux calibration of Gaia data, is among the largest of this kind to date
\citep{pancino12}. Observations started at the end of 2006, on six different
telescopes and instruments, and will presumably end in 2015. Almost 100\,000 raw
frames were collected and we faced the challenge of analyzing these data as
uniformely, automatedly, and carefully as possible. To ensure that the maximum
quality could be obtained from SPSS observations, a set of careful observations
\citep{pancino08,pancino09,pancino11} and data reduction
\citep{marinoni12,cocozza13} protocols was implemented, following an initial
assessment \citep{marinoni11,marinoni13,altavilla11,altavilla14} of all the
instrumental effects that can have an impact on the flux calibration precision
and accuracy. Our final goal was to make sure that all the instrumental effects
that could affect the quality of the final SPSS flux tables were under control,
and corrected with residuals below 1\%.

This paper presents the methods and results of such instrumental effects study,
and is organized as follows: Section~\ref{sec:shutter} presents the shutter
characterization of the employed CCD cameras; Section~\ref{sec:linearity} presents the
CCD linearity studies; Section~\ref{sec:calibs} presents the methods and results
of the calibration frames monitoring campaign; Section~\ref{sec:flex} presents a
study of lamp flexures with DOLORES at the TNG (Telescopio Nazionale Galileo);
Section~\ref{sec:polarization} presents a study of the effect of polarization on
the accuracy of flux calibrations; Section~\ref{sec:second} presents a method to
correct low resolution spectra for second-order contamination effects; and
finally Section~\ref{sec:concl} presents our summary and conclusions

\section{Shutter characterization}
\label{sec:shutter}

\begin{figure}
\centering
\includegraphics[width=\columnwidth]{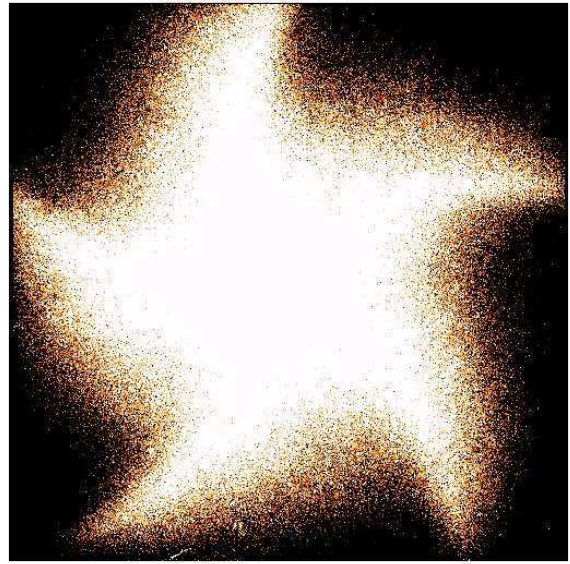} 
\caption{An example of non-uniform CCD illumination caused by a too short
t$_{\rm{exp}}$, for an iris-type shutter, obtained with LaRuca@1.5m in San Pedro
M\'artir, Mexico. The figure shows the ratio between a flat field obtained with
a relatively long exposure time (20~sec) and a very short one (0.1~sec): ADU
(Analogue-to-Digital Unit) variations of about $\pm$5\% are observed from the
center to the corners of the image.}
\label{fig:iris}
\end{figure}

To obtain accurate and precise photometry in our survey, we need to characterize
two shutter quantities: the {\em shutter delay} and the {\em minimum exposure
time}. 

The shutter delay time, also called shutter dead time or shutter offset, is the
difference between the requested and the effective t$_{\rm{exp}}$ (exposure time).
If the shutter closes slower than it opens, the exposure time will be longer than
requested, and the shutter delay might be positive, otherwise it is negative.
Knowledge of the shutter delay time is necessary for high-precision photometry,
where the effective t$_{\rm{exp}}$ is needed. 

\begin{table*}
\begin{center}
\caption{Flat field sequences used for the shutter delay, minimum acceptable
exposure time, and linearity measurements. The columns contain (see text for
more information): {\em (1)} the used instrument@telescope combination; {\em
(2)} the sequence of exposure times used; {\em (3)} the exposure time of the
monitoring exposures, when available; {\em (4)} the used filter or grism;
{\em(5)} the used slit width in arcseconds, when applicable; {\em (6)} the
observation date; {\em (7)} the experiment for which the measurements were used,
with {\em `a'} indicating the shutter delay measurement, {\em `b'} the minimum
acceptable exposure time measurement, and {\em `c'} the linearity measurement.
Of type $a$ and $c$ observations, the ones taken on 2009 January 29 lack
monitoring triplets, because the data were downloaded from the TNG archive.}
\begin{tabular}{l l r l c l l}
\hline
Instrument     & t$_{\rm{exp}}$                                       & t$_{\rm{exp}}^{\rm{m}}$ & Filter/grism & Slit & Date       & Notes\\
               & (sec)                                                & (sec)                   &              &  (") &   (yyyy Month dd)  &      \\
\hline                                                                           
BFOSC@Cassini  & 0.5, 1, 2, 3, 4, 7, 10, 20, 30, 40, 50, 55, 60, 65   & 5  & B     & --& 2009 May  27 & a,b \\
               & 6, 26                                                & 10 & gr7   & 2 & 2010 Aug. 30 & c   \\
EFOSC2@NTT     & 10, 60, 120                                          & 40 & gr3   & 2 & 2008 Nov. 26 & c   \\
               & 3, 15, 30                                            & 10 & gr3   & 2 & 2008 Nov. 27 & c   \\ 
               & 0.1, 0.2, 0.5, 1, 2, 3, 6, 15, 20                    & 10 & B     & --& 2008 Nov. 28 & a,b \\
DOLORES@TNG    & 5, 20, 55                                            & 25 & VHR-V & 2 & 2008 Jan. 17 & c   \\
               & 1, 2, 4, 5, 6, 7, 8, 9                               & -- & B     & --& 2008 Jan. 29 & a,b \\
CAFOS@2.2m     & 1, 5, 10, 20, 25, 30                                 & 15 & B200  & 2 & 2007 Apr. 01 & c   \\
               & 0.1, 0.5, 1, 2, 3, 4, 5, 6, 20, 30, 40               & -- & R     & --& 2008 Apr. 17 & b   \\
               & 0.1, 0.2, 0.5, 1, 2, 3, 4, 5, 6, 15, 20, 30          & -- & R     & --& 2008 Sep. 09 & b   \\
               & 0.1, 0.2, 0.4, 0.6, 0.8, 1, 4, 6, 8, 8.7             & 2  & V     & --& 2010 Sep. 23 & a,b,c \\
LaRuca@1.5m    & 0.1, 0.2, 0.4, 0.6, 0.8, 1, 2, 3, 4, 10, 20          & -- & White & --& 2008 Aug. 20 & b   \\
               & 1, 2, 4, 8, 15, 18, 20, 25, 30, 35, 36, 37           & 10 & B     & --& 2008 Aug. 23 & a,c \\
               & 2, 4, 7, 14, 28, 40, 56, 62, 68, 74, 80, 87, 89      & 20 & R     & --& 2010 Jul. 17 & a,c \\
               & 0.2, 0.4, 0.6, 0.8, 1, 2, 3, 4, 5, 6, 10, 15, 20, 30 & -- & R     & --& 2010 Jul. 17 & b   \\
\hline
\label{tab:series}
\end{tabular}
\end{center}
\end{table*}

The {\em minimum ac\-cep\-ta\-ble t$_{\rm{exp}}$} is set by the finite time the
shutter takes to travel from fully closed to fully open (and vice versa). It is
not strictly related to the shutter delay. A shutter can have a very small
shutter delay (for example because it opens and closes with very similar
delays), but it might take some time to completely open. A different shutter can
have a significant shutter delay (for example because it starts closing with a
significant delay), but it might have a negligible minimum acceptable time
because once it starts opening (or closing) it completes the operation in a very
short time. Regardless of the shutter delay, a shutter that takes a long time to
fully open (or close) might introduce significant\footnote{Owing to our
requirements for the flux calibration of Gaia data  \citep{FVL-001}, we cannot
accept illumination non-uniformities above 1\% approximately.}  illumination
non-uniformities across the CCD. To avoid illumination non-uniformities,
t$_{\rm{exp}}$ must be longer than a safe minimum (see Figure~\ref{fig:iris} for
an example).

\subsection{Observations and data reductions}
\label{sec:obshut}

To measure shutter effects, series of imaging flat fields with
different exposure times were obtained at each telescope (see
Table~\ref{tab:series}). Each series consisted of triplets of flats with
increasing exposure time, from very short to very long (thus useful for
measuring CCD linearity as well, see Section~\ref{sec:linearity}). After each
triplet, a monitoring triplet with constant exposure time was also taken, to
monitor the lamp intensity constancy\footnote{The most common cause of lamp
instability is thermal drift, causing the intensity to increase for a given
exposure time as the lamp gets warmer.}, following the so-called {\em bracketed
repeat exposure method} (see Section~\ref{sec:linearity} for more details).

All flat field images were processed with IRAF\footnote{IRAF is the Image
Reduction and Analysis Facility, a general purpose software system for the
reduction and analysis of astronomical data. IRAF is written and supported by
the IRAF programming group at the National Optical Astronomy Observatories
(NOAO) in Tucson, Arizona. NOAO is operated by the Association of Universities
for Research in Astronomy (AURA), Inc. under cooperative agreement with the
National Science Foundation.} by correcting them for overscan (where applicable)
and subtracting a master bias. For all analyzed instruments the dark currents
turned out to be negligible (see Section~\ref{sec:calibs}) and no correction for
dark was applied. The median of images belonging to the same triplet was
computed to remove cosmic ray hits and to reduce the noise. 

\begin{figure}
\centering
\includegraphics[angle=270,width=\columnwidth]{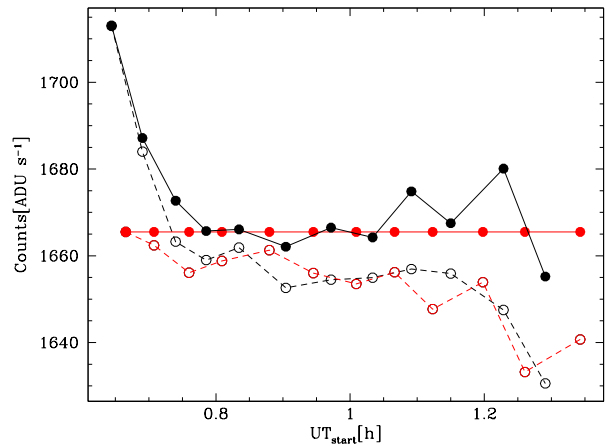} 
\caption{Example of lamp drift correction, based on data obtained with
LaRuca@1.5m in August 2008 (see Table~\ref{tab:series}). Median ADU per second
of each triplet are plotted as a function of time at triplet start. Red symbols
represent monitoring triplets, while black symbols represent the actual series
triplets. Empty symbols refer to uncorrected triplets, filled symbols to
triplets corrected for lamp drift (see Section~\ref{sec:obshut}). After
correction, monitoring triplets are aligned, while  data triplets present some
residual scatter (less than 1\%), presumably caused by residual lamp instability
after correction. A non-linear behaviour at low exposure times (left side of the
plot) is clearly visible.}
\label{fig:triplets}
\end{figure}

To correct for lamp drifts, each median image resulting from a monitoring
triplet was divided by a reference monitoring triplet (generally the first one
of the same sequence), to derive a correction factor. The correction factor was
used to correct each triplet --- both monitoring and data triplets --- and to
report them to an ideal ADU level, free of lamp drift variations. The results of
this correction are illustrated in Figure~\ref{fig:triplets}. The same type
of data sequences, extending to longer t$_{\rm{exp}}$, can be used for the CCD
linearity characterization described in Section~\ref{sec:linearity}, where the
lamp drift correction is performed in the same way.

\subsection{Shutter delay}

\begin{figure}
\centering
\includegraphics[width=\columnwidth]{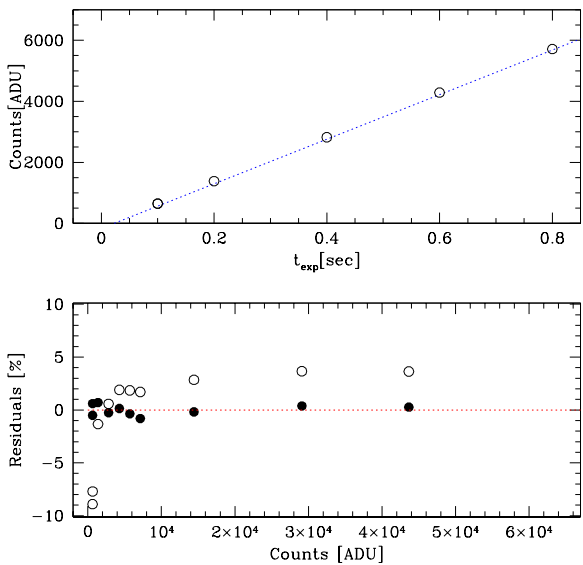} 
\caption{Example of shutter delay determination, for Calar Alto observations
obtained on 2010 September 23. {\em Top panel:} a zoom into the low
t$_{\rm{exp}}$ region of the classical ADU versus t$_{\rm{exp}}$ plot; the
dotted line represents a linear fit of all data and it does not intersect the X
axis at zero, but at --0.023~sec. {\em Bottom panel:} residuals of the count
rate (ADU s$^{-1}$) from their average value, as a function of ADU; empty dots
are the measured values, while filled dots are corrected for a shutter delay of
--0.011~sec. The average of the two determinations, --0.017~sec, is reported in 
Table~\ref{tab:shutter}.}
\label{fig:shutter}
\end{figure}

We used two different methods to measure the shutter delay. In the first method,
a linear extrapolation of the ADU (Analogue-to-Digital Units) versus
t$_{\rm{exp}}$ relation crosses the time axis at a value that is generally
different from zero, and that corresponds to the shutter delay
(Figure~\ref{fig:shutter}, top panel). In the second method, the count rate (ADU
per second) are plotted versus ADU and should ideally remain constant; in
practice, a deviation occurs at low t$_{\rm{exp}}$. By iteratively adjusting the
exposure time to $t+\delta t$, the shutter delay corresponds to the $\delta$t
that minimizes the residuals from a flat count rate, at the low t$_{\rm{exp}}$
end (see also Figure~\ref{fig:shutter}, bottom panel). We applied both methods
to the data in Table~\ref{tab:series}\footnote{We note that the BFOSC shutter
was changed in 2010 February, with a much faster one providing a negligible
shutter delay, but vignetting with t$_{\rm{exp}}$ lower than 5~sec (R.~Gualandi,
2010, private communication). Similarly, the old diaphragm shutter of CAFOS@2.2m
was replaced by a more efficient 2-blades shutter during Summer 2008 (Santos
Pedraz, 2010, private communication).}. We obtained consistent values and thus
we averaged the results from the two methods to obtain a final shutter
delay\footnote{In practice, we applied the linearity correction  (see
Section~\ref{sec:linearity}) to the shutter data and the shutter delay
correction to the linearity data, before re-computing both the linearity fit and
the shutter delay. As expected, such corrections had a negligible impact on the
resulting values.}. Results are reported in Table~\ref{tab:shutter}.

\subsection{Minimum acceptable exposure time}

The effect of illumination variations was estimated by dividing each of the flat
fields in a series (Table~\ref{tab:series}) by the longest non-saturated flat
field in the same series. The resulting ratio images (an example is shown in
Figure~\ref{fig:iris}) should have oscillations or variations which are caused
by noise only, and no systematic large scale patterns with amplitudes higher
than roughly 1\%, as required by the Gaia mission calibration goals. The minimum
acceptable exposure times, t$_{\rm{exp}}^{\rm{min}}$, obtained with this
criterion are listed in Table~\ref{tab:shutter}. While the shutter delay is
generally much lower than one second, t$_{\rm{exp}}^{\rm{min}}$ can be higher
than that. We stress the fact that if flat fields acquired with too short
t$_{\rm{exp}}$ are used to correct all frames of one night, they will affect
also scientific images taken with relatively long t$_{\rm{exp}}$, and should
thus be avoided.

\begin{table}
\begin{center}
\caption{Resulting shutter delays and minimum exposure times for the relevant
flat field series in Table~\ref{tab:series}. The columns contain (see text for
more information): {\em (1)} the used instrument@telescope combination; {\em
(2)}, the shutter delay; and {\em (3)} the minimum acceptable exposure time.}
\begin{tabular}{l l r }
\hline
Instrument       & $\delta$t               & t$_{\rm{exp}}^{\rm{min}}$ \\
                 & (sec)                   & (sec)                     \\
\hline                                                                           
BFOSC@Cassini$^{a}$ & --0.300$\pm$0.050    & $\simeq$5  \\
BFOSC@Cassini$^{b}$ & negligible           & $\simeq$5  \\

EFOSC2@NTT       &  +0.008$\pm$0.001       & $<$0.1 (if any) \\
DOLORES@TNG      & --0.011$\pm$0.002       & $<$1 (if any) \\
CAFOS@2.2m$^{c}$ & {\em (no data)}         & 3 \\
CAFOS@2.2m$^{d}$ & --0.017$\pm$0.070       &  $<$0.5 \\
LaRuca@1.5m      &  +0.028$\pm$0.004       & 5 \\
\hline
\multicolumn{3}{l}{$^{a}$Valid before 2010 February  (see text).}\\
\multicolumn{3}{l}{$^{b}$Valid after 2010 February (see text).}\\
\multicolumn{3}{l}{$^{c}$Valid before Summer 2008 (see text).}\\
\multicolumn{3}{l}{$^{d}$Valid after Summer 2008 (see text).}\\
\label{tab:shutter}
\end{tabular}
\end{center}
\end{table}

\section{CCD linearity}
\label{sec:linearity}

Linearity is a measure of how consistently the CCD responds to different light
intensity over its dynamic range. CCDs can exhibit non-linearities, typically at
either or both low and high signal levels \citep{djorgovski89,walker93}. High
quality CCDs show significant response deviations ($\geq 1$\%) from linearity
only close to saturation, i.e., at high signal  levels when the potential well
depth is almost full. Of course the CCD response strongly deviates at
saturation, when the potential well depth is full and additional incoming
photons do not increase the photoelectrons in a given pixel. When observations
are restricted to the linear portion of the dynamic range, the CCD performs as a
detector suitable for accurate spectrophotometric measurements. 

A few different methods have been presented in the literature to measure and
correct for CCD linearity losses, as the {\em bracketed method} described by
\citet{gilliland93}, the {\em bracketed repeat-exposure method}, and the {\em
ratio method} described by \citet{baldry99}. Another method related to the {\em
ratio method} is described by \citet{leach80} and \citet{leach87}: the {\em
variance method}. In our Gaia SPSS observing campaigns, we tested the CCD
linearity with two methods, described below.

\begin{figure}
\centering
\includegraphics[width=\columnwidth]{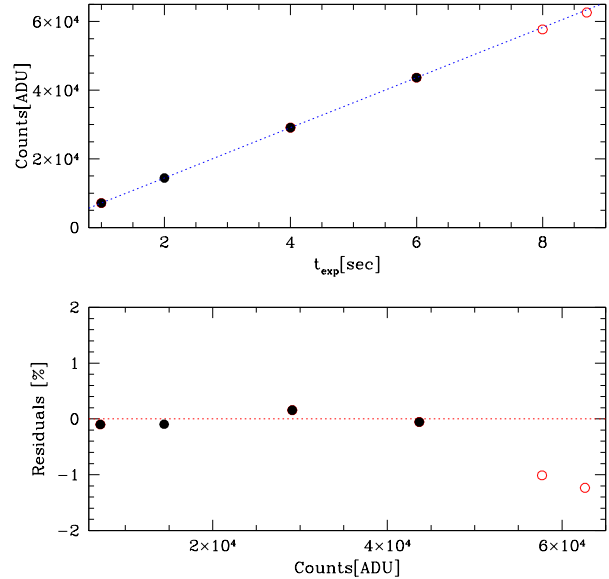} 
\caption{Example of linearity test for Calar Alto observations obtained on 
2010 September 23, with the SITE\#1d\_15 CCD. {\em Top panel:} a zoom into the high
ADUs region of the classical ADU vs. t$_{\rm{exp}}$ plot; the dotted line
represents a linear fit of all data. {\em Bottom panel:} residuals of the count
rate (ADU $^{-1}$) from their average value, as a function of counts; the dotted
line represents a constant count rate. In both panels, points deviating by more
than 1\% are represented by empty circles and coloured in red.}
\label{fig:lin1}
\end{figure}

The {\em classical method} is based on a series of imaging flat fields with
increasing t$_{\rm{exp}}$. Usually, at very high count levels the 1:1
relationship between ADU and t$_{\rm{exp}}$ breaks, and an increment of
t$_{\rm{exp}}$ does not correspond anymore to a fixed increase in ADU (an
example can be found in Figure~\ref{fig:lin1}). We used the relevant data listed
in Table~\ref{tab:series}, reduced and corrected for lamp drift as described in
Section~\ref{sec:obshut}. We considered that a deviation of $\geq$1\% from
linearity was not acceptable given our requirements, and thus we report in
Table~\ref{tab:lin} the ADU level at which such deviations occur. 


\begin{table*}
\begin{center}
\caption{Resulting maximum acceptable ADU levels obtained from the flat series
in Table~\ref{tab:series}. The columns contain (see text for more information):
{\em (1)} the used instrument@telescope combination; {\em (2)} the CCD used;
{\em (3)} the validity period of each CCD; {\em (4)} the maximum acceptable ADU
level obtained with the classical method; {\em (5)} the maximum acceptable ADU
level obtained with the \citet{stello06} method, when applicable; and {\em (6)}
the final adopted value for each CCD: when there were no sufficient data, the
safe value was assumed conservatively on the basis of that obtained for similar
CCDs; in those cases, the values are indicated in  parentheses.}
\tabcolsep=0.19cm
\begin{tabular}{l l l  c c c}
\hline
Instrument         & CCD        & Validity period  & ADU$_{\rm{max}}^{\rm{(Classic)}}$ & ADU$_{\rm{max}}^{\rm{(Stello)}}$ &  ADU$_{\rm{max}}^{\rm{(Adopted)}}$\\
\hline                                                                           
BFOSC@Cassini               & EEV 1300$\times$1340B (old) & Before 2008 Jul. 17    & ...     & ...     & (50\,000) \\
                            & EEV 1330$\times$1340B (new) & After  2008 Jul. 17    & ...     & 60\,000 &  60\,000  \\
EFOSC2@NTT (1$\times$1 bin) & CCD\#40 LORAL/LESSER        & All runs (from 2007)   & ...     & 49\,000 &  49\,000  \\
EFOSC2@NTT (2$\times$2 bin) &                             & All runs (from 2007)   & ...     & 47\,000 &  47\,000  \\
DOLORES@TNG                 & E2V 4240 (Marconi)          & All runs (from 2007)   & 60\,000 & 60\,000 &  60\,000  \\
CAFOS@2.2m                  & SITE\#1d\_15                & All runs (from 2007)   & 55\,000 & 53\,000 &  53\,000  \\
LaRuca@1.5m                 & SITE\#1                     & Until  2009 Jul.       & 60\,000 & ...     &  60\,000  \\
                            & ESOPO                       & 2009 Oct. 20--22       & ...     & ...     & (60\,000) \\
                            & E2V 4240 (Marconi) \#1      & 2009 Oct. -- 2010 Dec. & 60\,000 & ...     &  60\,000  \\
                            & E2V 4240 (Marconi) \#2      & 2011 Mar. 9--17        & ...     & ...     & (57\,000) \\
                            & SITE\#4                     & 2011 May  2--11        & ...     & ...     & (60\,000) \\
\hline
\label{tab:lin}
\end{tabular}
\end{center}
\end{table*}

\begin{table*}
\begin{center}
\caption{List of grisms used for the SPSS campaigns with the four
spectrographs. The columns contain: {\em (i)} the instrument@telescope
combination; {\em (ii)} the grism name; {\em (iii)} the minimum wavelength; {\em
(iv)} the maximum wavelength; {\em (v)} the approximate wavelength at which
second order contamination starts (if any, see text); {\em (vi)} the approximate
wavelength at which fringing starts (if any, see text); and {\em (vii)} the
approximate relative amplitude of the fringing pattern, as measured on our
spectra before correcting for fringing. All figures were derived from our SPSS
spectra.}
\begin{tabular}{l c c c c c c}
\hline
Instrument         & Grism  & $\lambda_{\rm{min}}$ & $\lambda_{\rm{max}}$ & $\lambda_{\rm{2nd order}}^{(start)}$ & $\lambda_{\rm{fringing}}^{(start)}$ & max fringing \\
                   &        & (\AA)                & (\AA)                & (\AA)                                &  (\AA)                              & ($\pm$\%) \\
\hline                                                                           
BFOSC@Cassini &  \#3 & 3300 &  6420 &          --- &          --- &    --- \\    
              &  \#5 & 4800 &  9800 &          --- & $\simeq$7000 & 10--15 \\    
CAFOS@2.2m    & B200 & 3200 &  9000 &          --- &          --- &    --- \\   
              & R200 & 6300 & 11000 &     $>$12000 & $\simeq$8500 & $\la$5 \\     
EFOSC2@NTT    &  \#5 & 5200 &  9350 &          --- & $\simeq$7300 &$\la$10 \\
              & \#11 & 3380 &  7520 &          --- &          --- &    --- \\    
              & \#16 & 6015 & 10320 & $\simeq$6000 & $\simeq$7300 &$\la$10 \\     
DOLORES@TNG   & LR-B & 3000 &  8430 & $\simeq$6000 &          --- &    --- \\    
              & LR-R & 4470 & 10073 & $\simeq$9500 & $\simeq$8000 &  5--15 \\     
\hline
\label{tab:grisms}
\end{tabular}
\end{center}
\end{table*}

\begin{figure}
\centering
\includegraphics[width=\columnwidth]{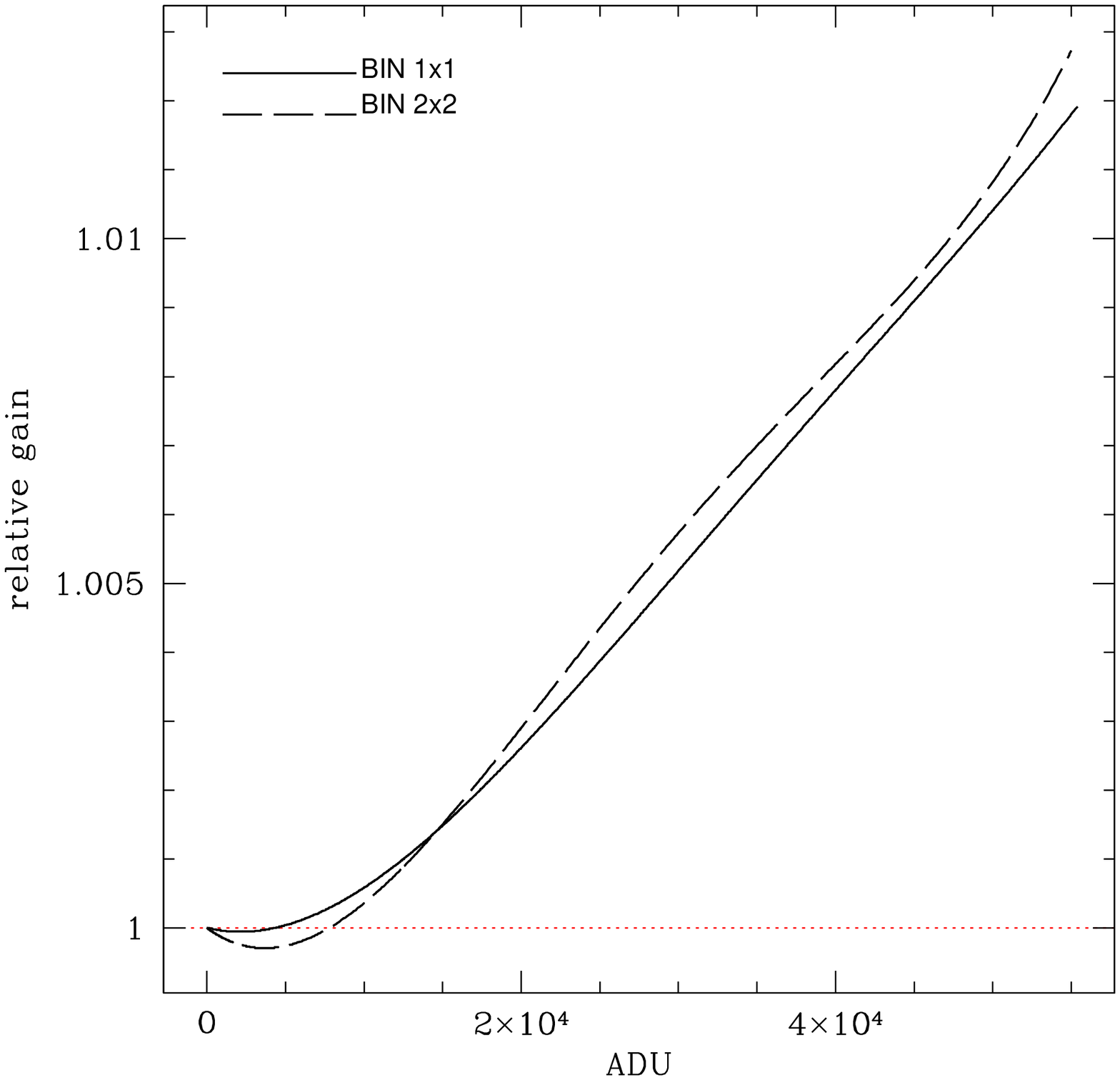} 
\caption{Final gain curves obtained with the \citet{stello06} method for NTT,
using data obtained on the 26 and 2008 November 27, with 1$\times$1 and
2$\times$2 binning, respectively, with the ESO grism \#3. The dashed line is for
the 2$\times$2 binning while the solid line is for the 1$\times$1 binning. Very
small deviations from linearity start to appear already around 5000~ADU, while
deviations above 1\% kick in at $\simeq$47\,500~ADU for the 2$\times$2 binning 
and at 49\,000~ADU for the 1$\times$1 binning.}
\label{fig:lin2}
\end{figure}

The second method we tested is described by \citet{stello06}, and is a variation
on the literature methods described above. It requires the acquisition of at
least two spectroscopic flat fields, one reaching the saturation level and the
other covering a fainter intensity range. The choice of slit width and grism (or
grating) is not crucial, but it is important to obtain flat fields with a wide
--- and preferably monotonic increasing --- intensity range along one direction
(either a CCD column or line). The frames were collapsed along the dispersion
direction, and corrected for lamp drifts as explained in
Section~\ref{sec:obshut}, but taking into account that lamp drifts in spectra
can show also colour variations. Afterwards, we applied the method as described
in the original paper by \citet{stello06}, by computing the intensity ratio of
exposure pairs, normalized by their respective t$_{\rm{exp}}$, which is called
{\em gain-ratio curve}. This ratio is the starting point of an iterative
procedure of inversion to determine the gain at different intensity levels. 

In conclusion, given our simple purpose of estimating a safe ADU limit for our
observations, the classical method and the high-precision method
\citep{stello06} give substantially the same results. While the required
spectroscopic observations are much faster to obtain, the \citet{stello06}
method is highly sensitive to different choices for the fit of the gain-ratio
curves (function, order, and rejection method), and to their extrapolation at
zero counts. Therefore, the method is not as straightforward to apply as the
classical method. To estimate the ADU level at which our data deviate more than
1\% from linearity, we conservatively chose the lowest of the two estimates, in
the few cases where the two methods provided different results. Results for the
employed instruments can be found in Table~\ref{tab:lin}.

\section{Calibration frames}
\label{sec:calibs}

The quality and stability of the calibration frames was tested on all the
acquired calibration data, producing recommendations for the calibration plans
that were progressively refined as more and more data were available for the
analysis. We found a surprising variety of behaviours among different
instruments and observing sites. Considering that we acquired of the order of
100\,000 frames in our campaigns \citep[see][for the SPSS campaigns
description]{pancino12}, it was important to implement automated procedures to
identify those calibration frames that needed a closer inspection, before
validating the corresponding reduced data. Whenever possible, we acquired daily
or nightly calibration frames, but in case of problems, we used the derived
calibration plan presented in Table~\ref{tab:plan} to decide how to analyze the
data (i.e., which of the archival or available frames were acceptable) and how
to judge the quality of the applied reductions. In the  following sections we
briefly describe the monitoring strategy and the main conclusions on each
instrument and telescope.

\subsection{CCD cosmetics}

We employed BPMs (Bad Pixel Masks), i.e., images of the same dimension of the
CCD frames, where good pixels are identified by zero values and bad pixels by
non-zero values. BPMs were created once or twice a year, using the ratio of two
flat fields with different count levels --- where pixels responding non-linearly
stand clearly out --- by means of the IRAF {\em ccdmask} task. They were applied
with the IRAF task {\em fixpix}, which replaces bad pixels with estimates of
their correct value. The estimates are built using a linear interpolation across
the nearest non-bad pixels, either across lines or columns, or both, depending
on the local characteristics of the BPM (see task description for more details),
i.e., differently for bad lines, bad columns, or clusters of bad pixels.

\begin{figure}
\centering
\includegraphics[width=\columnwidth]{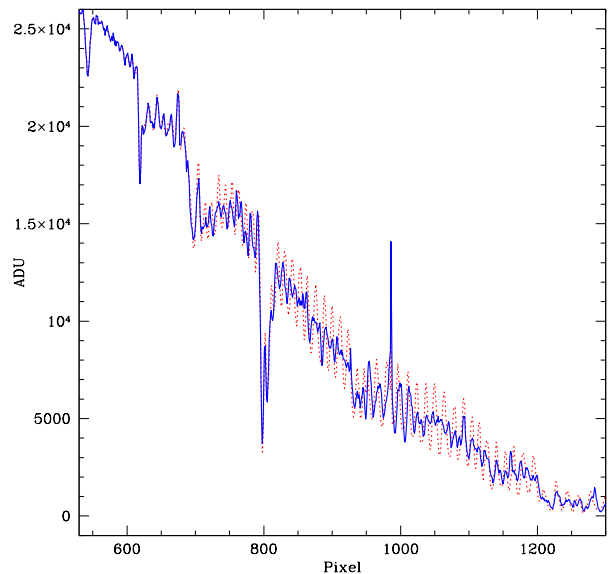} 
\caption{Example of fringing correction (see text) on a spectrum of  BD+284211,
observed in Loiano on 2013 October 28, with gr5 and a 2.5" slit. The blue solid
line is the spectrum after fringing correction, based on a day-time flat,
while the red dotted spectrum is before the correction. Beatings are observed,
caused by small differences in the fringing pattern of the spectrum and the flat
used for the correction. For example, aroud pixel 900 the residuals are below
5\%, while around pixel 1000 they remain higher, around 15\%.}
\label{fig:fringe}
\end{figure}

\subsection{Fringing}
 
Fringing, caused by multiple reflections and interferences of red or infrared
light in thin CCD substrates \citep{lesser90,howell12}, was not
relevant\footnote{As in the above cases, fringing is considered to be relevant
when the fringing pattern in images or spectra has an amplitude larger than 1\%
of the signal.} for the imaging filters used in our observations (mostly B, V,
and R)\footnote{The procedure for correcting fringing in images involves the
construction of a {\em super-flat} and is well described elsewhere
\citep{tyson90,newberry91,gullixson92}.}, but was present and relevant in all
our spectra, because we needed to cover the whole Gaia wavelength range, from
330 to 1050~nm. All the instruments used for spectroscopy do suffer from
fringing (see Table~\ref{tab:grisms}), well visible in our spectra (see
Figure~\ref{fig:fringe}). For NTT we took flat fields very close to our red
grism observations, while for all other telescopes we only relied upon day-time
lamp flats. However, while flat fielding does reduce the intensity of fringes,
the best fringing reduction strategy for spectroscopy, i.e., observing triplets
of spectra with the star in different positions along the slit, could not be
applied at the chosen telescopes because it would have been too time consuming,
observationally. 

The solution we adopted was to apply specific spectra processing steps to reduce
the impact of fringing, adapting the procedure used in the {\em
mkfringeflatcor}\footnote{http://stsdas.stsci.edu/cgi-bin/gethelp.cgi?mkfringef}
STIS (Space Telescope Imaging Spectrograph) task for IRAF \citep{malumuth03}.
Briefly, the most appropriate flat field available for each scientific spectrum
was extracted from a region covering exactly the same pixels covered by the
scientific spectrum, collapsed to a 1D spectrum along the dispersion direction,
and normalized to one. The region that does not contain a significant fringing
pattern was then flattened exactly to 1, to avoid adding extra noise to the
scientific spectra. Finally, this specially prepared 1D spectroscopic flat was
aligned to the scientific spectra and scaled until the residuals of the fringing
pattern were minimized, using the IRAF {\em telluric} task, and applied to
scientific spectra to correct them for fringing. 

We found that even using spectroscopic flat fields obtained in the day-time
(thus with different incident light patterns), the procedure greatly helps in
reducing the amplitude of fringes, up to a factor of 2--3 in relative intensity
(see Figure~\ref{fig:fringe}). Sometimes, the closeness in position of the
day-time flats fringe patterns to the SPSS fringe pattern produced beatings in
the residuals of the fringing correction, with regions were the pattern was
almost completely erased and regions where a residual fringing pattern remained.
This has to be taken into account for applications of the method to scientific
cases that could be sensitive to this kind of residual effects.

Additionally, in our spectroscopic campaign we often have observations of the
same SPSS from different telescope and observing nights, and thus spectroscopic
fringing residuals can be further reduced by building a median of the different
spectra. The very red end of spectra ($\lambda\ga$9500~\AA), where fringing is
strong and S/N is low, will also be replaced with model or semi-empirical
template spectra, as done for example by \citet{bohlin07}.

\subsection{Dark frames}

\begin{figure}
\centering
\includegraphics[angle=270,width=\columnwidth]{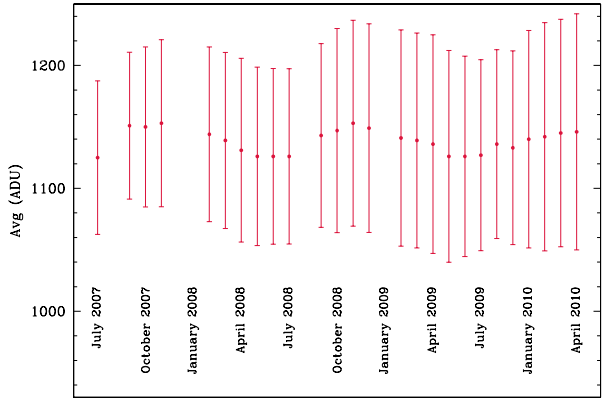} 
\caption{Example of long-term behaviour of the average counts (with their sigma)
of REM monthly master dark frames with t$_{\rm{exp}}$=60~s, covering roughly 3 years. A
seasonal trend is clearly visible.}
\label{fig:dark}
\end{figure}

Dark correction is the subtraction of the electron counts which accumulate in
each pixel due to thermal noise. The reduction of dark currents is the main
reason why all astronomical CCDs are cooled to liquid nitrogen temperatures. We
took dark exposures as long as the longest scientific exposure, at least once
per year. As expected, all used instruments showed at most a few ADU per pixel
per hour, except for the REM (Rapid Eye Movement) robotic telescope in La Silla,
which is cooled by a Peltier system. The REM staff creates monthly master dark
frames, containing the bias structure as well (see below), thus we applied the
closest master dark frame in time, taken with the correct t$_{\rm{exp}}$. A
seasonal effect on the dark frames average level can be clearly seen in
Figure~\ref{fig:dark}, while the 2D shape of the frames was extremely stable and
the ADU increased linearly with t$_{\rm{exp}}$.

\begin{table*}
\begin{center}
\caption{Calibration plan recommendations following our calibration frames
monitoring (see text). For each type of calibration frame, the minimum
acceptable stability timescale is reported. }
\begin{tabular}{l c c c c c c c}
\hline
Instrument    & Dark    & Photometric & Spectroscopic & Photometric & Photometric & Spectroscopic & Spectroscopic \\
              &         & Bias        & Bias          &   Dome Flat & Sky Flat    & Lamp flat     & Sky flat      \\
\hline                                                                           
BFOSC@Cassini &     --- & 4 days      & 4 days        & 300 days    &varies$^{c}$ & 1 day         & 1 week \\   
EFOSC2@NTT    &     --- & 5 days      & 1 day         & 250 days    & 250 days    & 1 day         & 1 week \\ 
DOLORES@TNG   &     --- & 1 day       & 1 day         & ---         & 250 days    & 5 days        & 1 week \\   
CAFOS@2.2m    &     --- & 1 week      & 1 week        &varies$^{b}$ & 1 day       & 1 day         & 1 week \\
LaRuca@1.5m   &     --- &varies$^{a}$ & ---           & ---         & 5 days      & ---           & ---    \\
ROSS@REM$^{d}$& 1 month & ---         & ---           & ---         & ---         & ---           & ---    \\
\hline
\multicolumn{8}{l}{$^{a}$Varies from 1 to 7 days, depending on the CCD used.}\\
\multicolumn{8}{l}{$^{b}$Was 1 day until Summer 2008, and 1 week afterwards.}\\
\multicolumn{8}{l}{$^{c}$Was 4 days until 2010 February, and 100 days afterwards.}\\
\multicolumn{8}{l}{$^{d}$REM flat fields were not applied, to avoid reinforcing
the strong light-concentration effect from which ROSS suffers.}\\
\label{tab:plan}
\end{tabular}
\end{center}
\end{table*}

\subsection{Overscan and bias frames}

\begin{figure}
\centering
\includegraphics[angle=270,width=\columnwidth,]{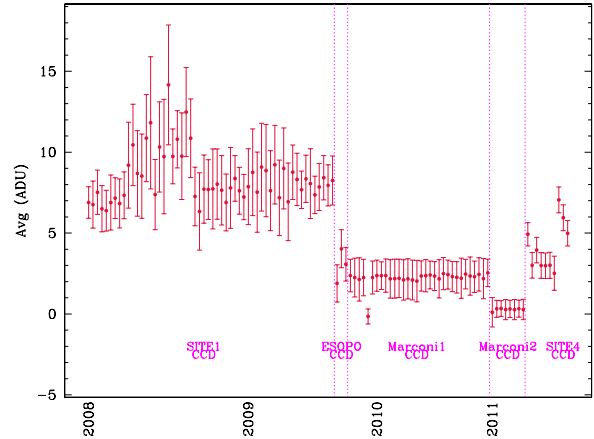} 
\caption{Example of long-term behaviour of the average counts (with their sigma)
of San Pedro M\'artir daily master bias frames, covering more than three years.
The CCD changes are marked and correspond to clear jumps in ADU and also in
variations of the ADU stability with time.}
\label{fig:bias}
\end{figure}

The bias current is an offset, preset electronically, to ensure that the
Analogue-to-Digital Converter (ADC) always receives a positive value and
operates in a linear regime, as much as possible. The offset for each exposure
given by the bias level has to be subtracted before further reduction, and may
be modelled as $A+B(x_i, y_i)$, where $B$ is the pixel-to-pixel variation or 2D
structure, which is time-invariant unless particular problems occur, while $A$
is the overall bias level, that can change slightly during the night or even
seasonally, owing to temperature changes. 

Generally $A$ can be measured using a strip of pixels, called {\em overscan},
acquired by continuing to readout the CCD beyond its real physical extent: the
result is an oversized array with a strip of signal-free pixels. In our survey,
the overscan was available only for DOLORES@TNG and LaRuca@1.5m in San Pedro
M\'artir, thus in all other cases we used the 2D bias frames to implicitly
correct for $A$ as well. This produces some additional uncertainty if $A$ varies
during the night\footnote{However, we found that in both the photometry and
long-slit spectroscopy cases, the standard sky subtraction procedures we adopted
were sufficiently accurate to get rid of any residual bias offset, at least
within the required 1\% level.}, because the bias calibration frames were
generally acquired in the afternoon (and also in the morning for the less stable
instruments, like LaRuca). The different stability of the four CCDs used with
LaRuca is illustrated in Figure~\ref{fig:bias}, where the bias level --- after
overscan subtraction\footnote{It can happen that after overscan correction the
average level of a bias frame remains above zero by a few ADUs, because of the
way the CCD is read out.} --- of all the master bias frames is plotted.

\subsection{Flat fields}

\begin{figure}
\centering
\includegraphics[angle=270,width=\columnwidth]{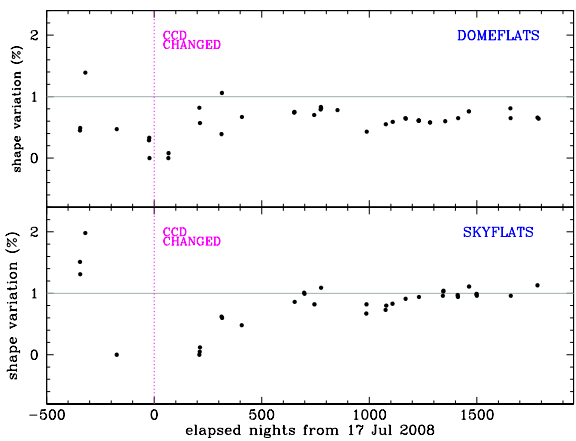} 
\caption{Example of long-term behaviour of the normalized shape variation
($\Delta S/S$, see text) of BFOSC nightly dome (top panel) and sky (bottom
panel) flat fields in B filter. The shape variations are rarely above 1\%,
globally, showing an impressive stability over a period covering more than 2000
nights. The CCD change in Summer 2008 is marked by a vertical dotted line.}
\label{fig:flat}
\end{figure}

After bias subtraction and --- if needed --- dark correction, data values are
directly related to the number of photons detected in each CCD pixel. But
different pixels can be characterized by a different sensitivity: the flat
fielding correction accounts for the non-uniform CCD response to incident light,
that can show variations on all scales, from the whole CCD to the single pixel.
In imaging, flat field frames are acquired using the twilight sky or a screen
uniformly illuminated by a lamp, with each filter; they are used to correct for
sensitivity variations on all scales, acting as a sort of {\em sensitivity map}.
\citet{massey13} recommend the use of dome or screen flat fields to correct for
the small-scale sensitivity variations and of sky flat fields for large-scale
variations. In spectroscopy, only the small scale variations are corrected with
flat fields obtained through the spectrograph, i.e., with a slit and a grism (or
grating). Of course, flat field frames with low signal or saturated were always 
rejected (see Section~\ref{sec:linearity}).

For imaging flat fields, we were interested in monitoring the large scale shape
variations with time. We thus took all the masterframes obtained for each
instrument, nor\-ma\-li\-zed them by their mode, and smoothed them with a boxcar
to remove small scale variations. We then computed a stability function
K$_{\rm{stab}}$, using the pixel-to-pixel difference of the counts in each
masterflat ($F_{\rm{pix}}$) with respect to a reference masterflat
($F^{\rm{ref}}_{\rm{pix}}$):
\begin{equation}
K_{\rm{stab}} =  \sum_{pix=1}^{n} (F_{\rm{pix}} -
F_{\rm{pix}}^{\rm{ref}})^2
\end{equation} 
We can then define a {\em normalized shape
variation} as 
\begin{equation}
\frac{\Delta S}{S} =
\sqrt{\frac{K_{\rm{stab}}}{K_{\rm{stab}}(1\%)}}
\end{equation} 
where $K_{\rm{stab}}(1\%)$ is
computed in the case where the pixel-to-pixel variation is equal to 1\%
everywhere. Using this indicator, we have a significant variation (i.e., a
global variation above 1\%) when $\Delta S/S>1$. 

Using the K$_{\rm{stab}}$ global indicator, we explored all the master flat
fields obtained during our SPSS survey with automatic procedures (see
Figure~\ref{fig:flat} for an example), thus identifying the flat fields that
needed to be visually inspected for anomalies, before applying them to the
scientific images.

\subsubsection{Illumination variations}

For imaging, dome and sky flat fields are sufficient to correct for the CCD
response variations \citep{massey97,massey13}. Instrumental effects, like
internal light reflections or light concentration, need instead ad hoc
procedures \citep{manfroid01,koch04}. Among the imagers used in the SPSS
campaigns, only ROSS@REM showed a significant (up to 5--10\%, depending on the
amount of incident light) cushion-shaped illumination variation, that was
worsened by the application of the flat-field correction, similarly to what
happens with light concentration. Because ROSS@REM was only used for relative
photometry, we decided not to apply any flat-field correction to REM images and
to keep the SPSS as much as possible in the same X and Y position in the CCD
during each time series. Nevertheless, relative photometry with ROSS@REM proved
to be extremely difficult and only rarely met our requirement of $\sim$millimag
accuracy.

For spectroscopy, illumination variations caused by imperfections on the slit
borders, vignetting, and other similar effects, are usually corrected with the
help of twilight sky spectroscopic flat fields \citep{massey13,schonebeck14},
applied after normal (i.e., lamp) flat fielding. After some initial testing, we
decided to obtain spectroscopic flats fields once per run, as a compromise
between the stability of the results and the time needed to obtain the flats
(see Table~\ref{tab:plan}).

\section{Lamp flexures}
\label{sec:flex}

Low resolution spectrographs are usually mounted at the telescope or at the
telescope derotator, and they move during observations. At each different
position, the varying projection of the gravitational force leads to mechanical
distortions, which can be seen on the wavelength calibration frames, where they
produce linear or non-linear shifts of the lamp emission lines \citep{munari92}.
A correct flux table must associate the right flux to the right wavelength, thus
any error in wavelength has an impact on the flux calibration of a
spectrum\footnote{It is important to remark that --- especially in those parts
of the spectrum where the flux changes rapidly with wavelength --- even small
errors in the wavelength calibration can imply relatively large errors on the
flux calibration. This problem becomes difficult to solve, i.e., by means of
cross-correlation to align spectra, for featureless or nearly featureless stars,
which are in general the best flux calibrators.}.

We performed a test to evaluate the DOLORES@TNG lamp flexures on 2008 January 31, 
when the instrument was equipped with three separate calibration lamps
(He, Ne, Ar) that could not be switched on simultaneously, and the procedure to
obtain high S/N calibration lamps for each single star during night-time
observations was too time consuming. We used the LR-R grism, the 2" slit and the
Ar lamp.  Triplets of wavelength calibration lamp spectra were acquired at
different positions of the derotator, covering a complete derotator circle, from
-260 to + 100 degrees in steps of 10 degrees forward, and from +95  up to -255
degrees in steps of 10 degrees backwards. The median of the three acquired lamp
frames at each position was used to evaluate line shifts.

\begin{figure}[t!]
\centering
\includegraphics[width=\columnwidth]{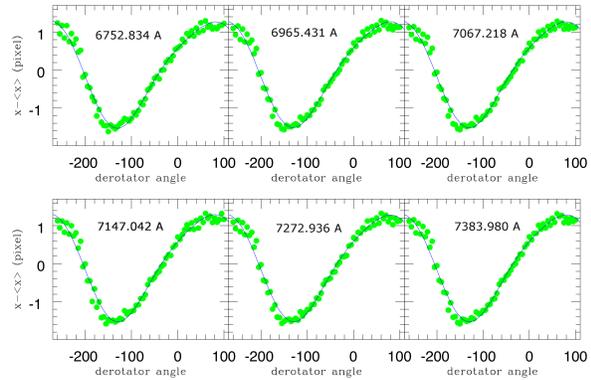} 
\caption{Example of lamp flexure effect on emission lines position of a TNG
Argon wavelength calibration lamp exposure with the LR-R grism (see text). Each
panel represents the line peak shift in pixels (ordinate) of six individual
argon lines as a function of DOLORES derotator angle position (abscissa). Each
green dot corresponds to a peak measurement in a different lamp exposure. The
typical shift of almost $\pm$1.5 pixels corresponds to almost 8~\AA. The
thin black line is a polynomial fit to the data, to guide the eye.}
\label{fig:lamp}
\end{figure}

The resulting shifts of emission lines, as a function of derotator angle, are
characterized by a quasi-sinusoidal trend, as shown in Figure~\ref{fig:lamp} for
a selection of the examined lines. The typical size of the global oscillation
was almost three pixels, with an error on the lines peak position ranging from
0.01 pixels in the central portion of the CCD to 0.06 pixels at the CCD borders.
The shift amounts to almost 8~\AA\  in the space of wavelengths and thus daytime
lamps are clearly not appropriate for the wavelength calibration of spectra in
the Gaia spectrophotometric campaigns. For this reason, most astronomers also
take wavelength calibration lamps close to the position of each observed
spectrum. Not being able to measure emission line shifts for all instruments, we
also adopted the same strategy. In particular, if the night lamp observations
required too much time --- as was initially the case with TNG --- we decided to
employ day-time lamps with very high S/N, and shift them to match the night
lamps taken with lower S/N in order to save time. In all other cases we used
night-time lamps.

\section{Polarization}
\label{sec:polarization}

The reflectivity of curved and plane mirrors may be dif\-fe\-rent for different
directions of linear polarization, in a wavelength dependent way \citep[see][and
references therein]{breckinridge04}. An incoming beam of non-polarized light
will be polarized according to this difference of efficiency. This might affect
the sensitivity of the whole observing system, especially if the light beam
encounters other polarization-sensitive optical elements after reflection. The
degree of induced polarization depends on the {\em angle of incidence}: the
maximum degree of (linear) polarization occurs for an incidence angle which
depends on the material, called {\it polarization angle}. Refractive dispersing
elements (grisms, in our case) may also --- in principle --- produce some degree
of polarization, or may have different transmission efficiencies for different
{\it intrinsic} polarization of the incoming light, thus introducing selective
light losses. A combination of optical elements with different orientation could
thus produce a significant variation in the efficiency, dependent on wavelength
and/or pointing.

\begin{figure*}[t!]
\centering
\includegraphics[width=\columnwidth]{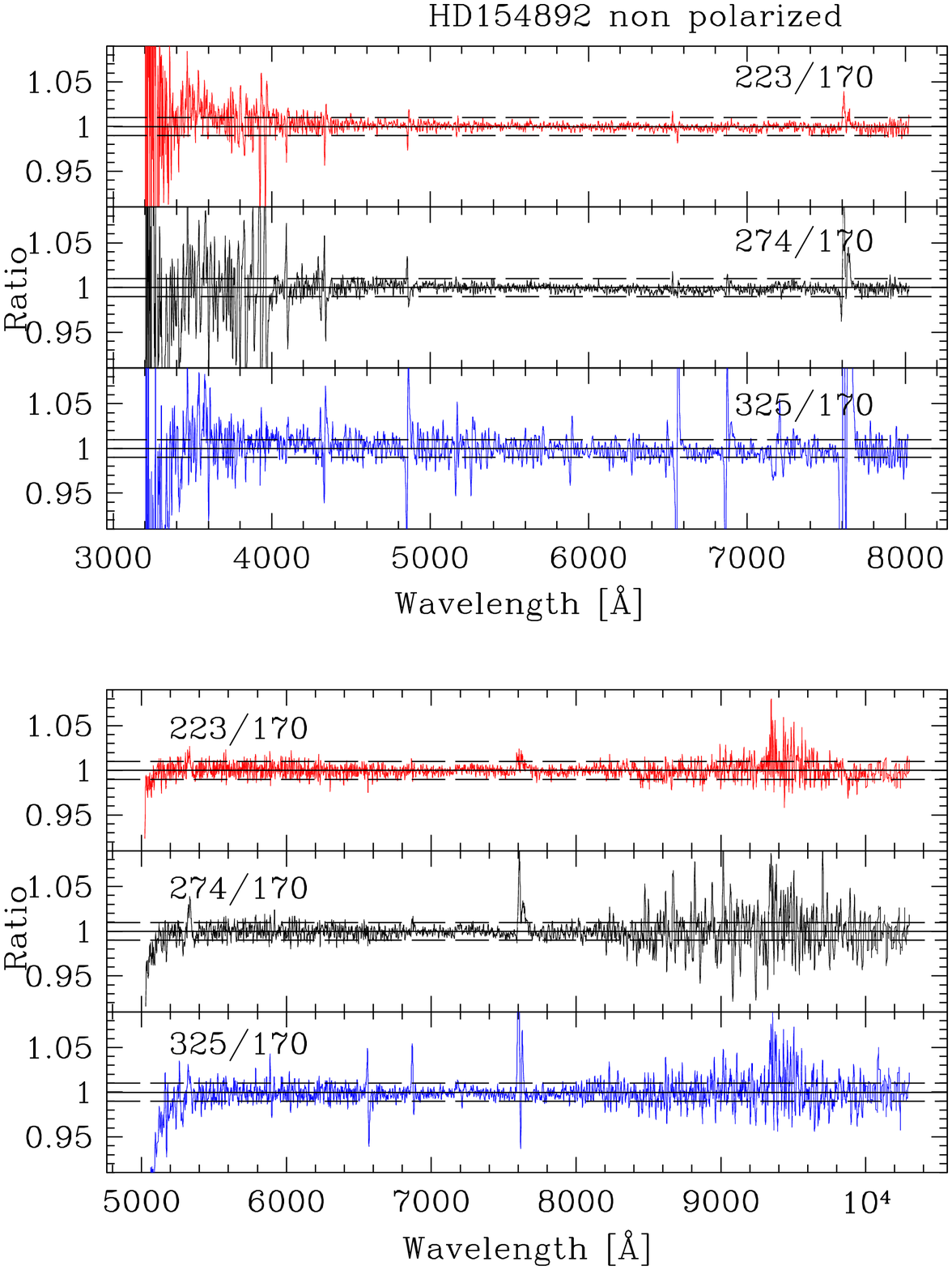} 
\includegraphics[width=\columnwidth]{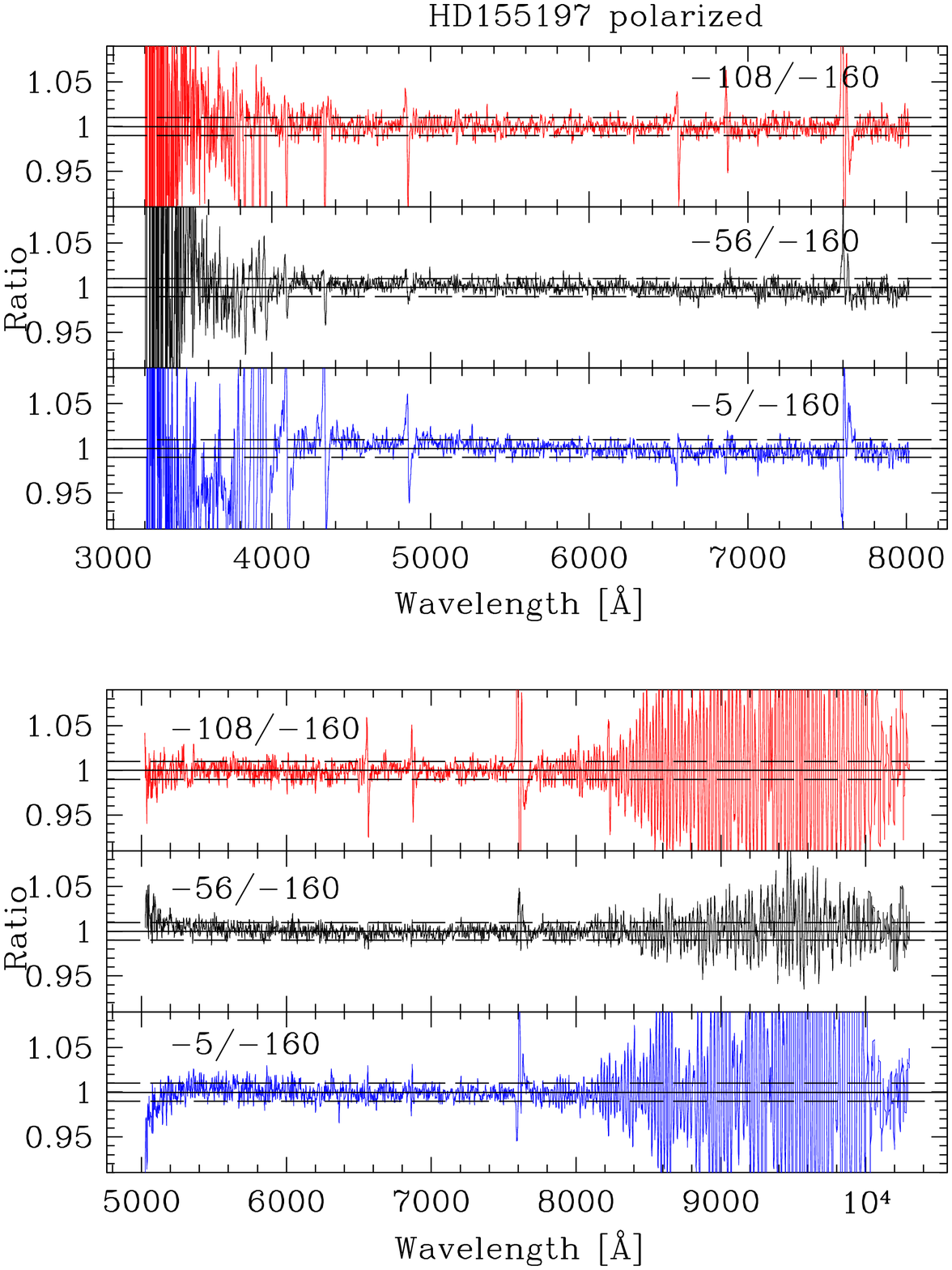} 
\caption{Results of our test on the effect of polarization on spectrophotometry.
The left panels report observations of the ``non-polarized" star HD~154892.
while the right panels of the ``polarized" star HD~155197 (see text for more
details). The three top panels for each star refer to the LR-B grism, while the
bottom panels to the LR-R grism. Each sub-panel shows the ratio of spectra taken
at different derotator angles with respect to a reference angle, which is
170~deg for the non-polarized star and --160~deg for the polarized star. All
ratios are shown as a function of wavelength. Dashed lines in each sub-panel
indicate $\pm$1\% ratio variations. Except for the noisy blue edge, and the red
parts were fringing is important, it is clear that the effect is negligible
for our purposes. The increase in the fringing residuals pattern for the
polarized star is caused by the fact that the polarized star was accidentally
placed in different positions along the slit for different polarization angles.}
\label{fig:polarization}
\end{figure*}

We can envisage three cases relevant to our SPSS campaigns:
\begin{enumerate}
\item{The light beam encounters just one optical element potentially sensitive 
to polarization, for example the grism, as in the case of  CAFOS@2.2m in Calar
Alto. If the incoming light is polarized, there may be a difference in the
response depending on the actual orientation of the grism in the plane
perpendicular to the direction of the light beam. If the incoming light is not
polarized, there is no effect.}
\item{There is more than one element potentially sensitive to polarization, but 
their relative orientation in the planes perpendicular to the direction of the 
incoming beam are fixed, as is the case of the flat mirror feeding the grisms
of BFOSC@Cassini. The net result is the same as above.}
\item{There is more than one element potentially sensitive to polarization, and
their relative orientation in the planes perpendicular to the direction of the
incoming beam changes, depending on the telescope pointing. This is the case of
the flat M3 mirror of the twin TNG and NTT telescopes\footnote{The EFOSC2@NTT
web site  (http://www.eso.org/sci/facili ties/lasilla/instruments/efosc-3p6/)
reports: {\em ``Preliminary analysis of instrument polarization shows $<$ 0.1\%
at field center, and $\simeq$ 0.4\% at the edge (07/07/2005)''.}} that drives
the main light path to the Nasmyth focus where it feeds the grism of the DOLORES
and EFOSC2 spectrographs, mounted on the derotator device. In this case some
polarization effect can be induced even on non-polarized incoming light, and the
grism may respond slightly differently --- in terms of efficiency --- to
differently po\-la\-ri\-zed light, resulting in a light loss depending on the
relative orientation of the two elements \citep[see][]{giro03}.}
\end{enumerate}

\subsection{The TNG polarization experiment}
\label{sec:TNGpolarization}

While potentially highly polarized sources should be avoided as candidate SPSS
(as for instance DP white  dwarfs or heavily extincted stars), there is no
observational strategy that can completely avoid the problem. We thus tested the
above hypothesis on TNG, because it has the same NTT design, and they are the
only two telescopes in our campaign where polarization could significantly
affect spectro-photometric observations. 

We observed two stars from \citet{turnshek90}: HD~155197, which exhibits a
polarization in V band of 4.38$\pm$0.03~\%  --- that we will call {\em polarized
star} hereafter --- and HD~154892, having a B-band polarization of
0.05$\pm$0.03~\% --- that we will call {\em non-polarized star}. They were
observed on a clear night (2012 June 1, kindly obtained by the TNG staff), with
good seeing (0.8"--1.5"), using a wide slit (10") and the same grisms used for the
SPSS campaign (LR-R and LR-B, see Table~\ref{tab:grisms}). Spectra were taken at
different derotator angles of --160, --108, --56, and --5 degrees for the
polarized star, and 170, 223, 274, and 325 degrees for the non-polarized star. The
spectra were bias and flat-field corrected, extracted, and calibrated in
wavelength with the help of narrow slit (2") spectra and day-time calibration
lamps, using standard recipes from the IRAF {\em apall} and {\em onedspec}
packages, configured exactly as done for our SPSS reductions \citep{cocozza13}. We
finally corrected all spectra for atmospheric extinction, using an extinction
curve built from our own data, even if the airmass difference among the used
spectra was small: 0.10 for the polarized star and 0.05 for the non-polarized
star.

As can be seen in Figure~\ref{fig:polarization}, the overall shapes of the
spectra taken at different derotator angles agree with one another within 1\%,
except at the blue edge (below 4000 \AA), where the S/N is rather low, and above
8000 \AA, as expected because of fringing and telluric absorption (which were
not corrected for in this particular test). This confirms first of all that the
night conditions were photometric. The additional absence of significant slopes
or oscillations confirms that both the stellar intrinsic polarization and the
polarization supposedly induced by the instrument configuration had no
significant impact on our flux measurements (i.e., not more than 1\%), even for
different stellar polarization levels, wavelength regions, or derotator
positions.

\section{Second-order contamination}
\label{sec:second}

Low resolution spectrographs and large format CCDs allow for observations of a
wide wavelength range, as needed for the Gaia SPSS grid. However, with both
grisms and gra\-tings the light from different orders can overlap, especially if
no cross disperser or blocking filter is employed \citep{gutierrez94}.
Typically, light from the blue wavelengths of the second order can contaminate
the red portion of the spectra, in the spectrographs used in our campaign. The
characteristics of the contaminating light depend on both the spectral energy
distribution of the observed source and the characteristics of the dispersing
element. In particular, it is necessary to know the grism \citep{traub90} or
grating \citep{szokoly04} equation to apply a correction. 

Two of our adopted instrument configurations, for EFOSC2@NTT and DOLORES@TNG, are
known to possess significant second order contamination. For CAFOS@2.2m in Calar
Alto the only visible second order light started well after 10\,000~\AA, after
the end of the first-order spectrum in the red\footnote{The manual mentions a
second order contamination for the B200 grism starting after 6\,500~\AA, but our
data reach up to $\simeq$8\,500~\AA\  and no contaminating light was detected;
for R200, the manual mentions a second-order contamination starting at
10\,000~\AA, but we could clearly see the second order spectrum, starting well
after the end of the first order spectrum, with no overlap between the two
orders.}. For BFOSC@Cassini it was absent, both judging from the BFOSC manuals
and web pages, and from the quality of our final spectra of
CALSPEC\footnote{http://www.stsci.edu/hst/observatory/crds/calspec.html} stars.
The chosen setups were unavoidable owing to our requirements to cover as much as
possible the Gaia wavelength range (300-1100~nm) and to oversample the
re\-so\-lu\-tion of the BP and RP (Blue and Red Photometers) by a factor of
up to 4--5, whenever possible. 

We thus employed an adaptation of the method by \citet{sanchez06} to correct our
spectra from second-order contamination. The original method requires
observations of a blue and a red star, with well known calibrated flux in the
literature. They are used to build a {\em response curve} for the contaminating
second-order spectra, by solving the following system of equations:
\begin{equation}
S_b=C_1T_b+C_2T_{2b}
\end{equation}
\begin{equation}
S_r=C_1T_r+C_2T_{2r}
\end{equation}
 where the $b$
subscript refers to the blue star and the $r$ one to the red star. $S$ is the
observed spectrum, including the second-order contamination, while $T$ is the
true spectrum obtained from the literature (thus free from second-order
contamination), and $T_2$ is the second-order contaminating component. $C_1$ is
the response curve for the first order light and $C_2$ for the second order
light: once they are known, any spectrum observed with the same instrument can
be corrected for second-order contamination as described by \citet{sanchez06}.
Since $C_1$ and $C_2$ were not computed each night in our SPSS campaign, we
derived the corrected spectrum ($S^{\prime}_{b}$ for the blue star case) using
\begin{equation}
S^{\prime}_b=S_b-C_2 S^{\prime}_{2b}
\end{equation}
 where $S^{\prime}_{2b}$ is
the ratio $S_b/C_1$, i.e., the second order contaminating spectrum, after
resampling and shifting, according to the function mapping the first order into
the second order wavelength range (see following sections for more details), and
still not calibrated in flux. 

\begin{figure}
\centering
\includegraphics[angle=270,width=\columnwidth]{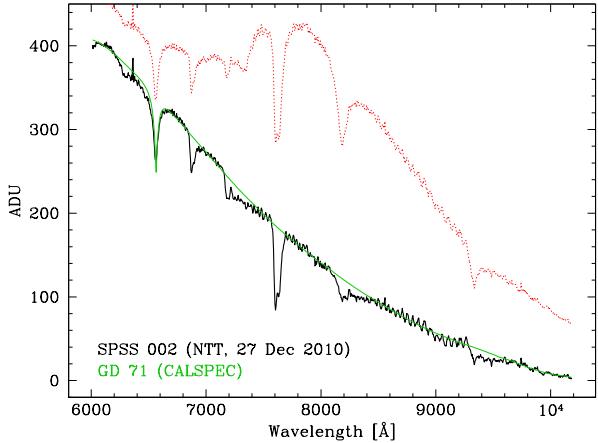} 
\caption{Example of our second order contamination correction for a CALSPEC star
(GD~71, one of the pillars) observed with NTT. The red (dotted) spectrum is our
observed spectrum contaminated by second-order light coming from blue
wavelengths. The black spectrum is the same spectrum after our second order
correction procedure; it still contains telluric absorption bands and fringing.
The green spectrum is the CALSPEC spectrum for the same star: the spectral shape
is recovered to the 1\% level.}
\label{fig:NTTsecond}
\end{figure}

The method by \citet{sanchez06} requires a single spectrum to cover the whole
wavelength range, while in our case we always had two setups --- with some
overlap --- observed in two slightly different moments of time. Our adaptation
consisted in joining the blue and red spectra of the observed stars, after
reporting the blue spectrum to the airmass of the red one with a good extinction
curve. A residual jump in the junction point does not have any effect on the
quality of the final correction for second order. 

While the original method by \citet{sanchez06} for deriving $C_1$ and $C_2$ can
be applied directly, even in non-perfectly photometric conditions, our
particular adaptation requires some care in those cases where extinction
vatiations are large. Application of the method is safe in photometric and
grey\footnote{Grey conditions are attained when the cloud coverage produced grey
extinction variations, i.e., when extinction does not alter significantly the
spectral shape (within 1\%). This condition is almost always verified in the
case of veils or thin clouds \citep{oke90,pakstiene03}, and can easily be
checked a posteriori.} nights, where extinction variations are lower than
$\simeq$3\%, roughly speaking.

\subsection{NTT second order correction}

The spectra of two well known flux standards, Feige~110 \citep[a blue
star,][]{landolt07} and LTT~1020 \citep[a red star,][]{landolt92}, where
acquired with the blue and red EFOSC2 grisms \#11 and \#16 (see
Table~\ref{tab:grisms}), used for the whole SPSS campaign. Each star was
observed with a 10'' slit to gather all the flux, and with a 2'' slit to obtain
a reliable wavelength calibration. 

The spectra were extracted and wavelength calibrated as described in
Section~\ref{sec:TNGpolarization}. We created an extinction curve from our own
data, that compared well with the CTIO and ESO ones. We then joined the grism \#11
and \#16, reported to the same airmass, into a single spectrum, cutting them at
6000~\AA, i.e., well before the expected start of second order contamination. We
used the absorption lines of the blue contaminating spectrum --- well visible at
red wavelengths --- to determine the function that maps the blue contaminating
wavelengths in the red wavelength domain: 
\begin{equation}
\lambda_2 = -3815+3.271~\lambda_1
-0.0001167~\lambda_1^2
\end{equation}
 where $\lambda_1$ is the wavelength in \AA\ 
of the first order, and $\lambda_2$ the corresponding wavelength of the
second-order contaminating blue light. The residuals around this fit were of
2.9~\AA. The flux reference tables used for computing the response curves, i.e.,
$T_b$ and T$_r$ in the equations of the previous section, were obtained for
LTT~1020 from CTIO\footnote{ftp://ftp.eso.org/pub/stecf/standards/ctiostan}
\citep{hamuy92,hamuy94} and for Feige~110 from the CALSPEC database
\citep{bohlin01}.

Once the two response curves $C_1$ and $C_2$ were derived, we computed the
percentage of the blue light of a star that is expected to fall at each red
wavelength, as 
\begin{equation}
P=\frac{C'_2}{C_1}
\end{equation}
 where $C'_2$ is $C_2$ reported
into blue wavelengths using the relation above. The computation of $P$ was
repeated in two different epochs (using LTT~1020 and Feige~110 on 2008 November
30, and LTT~9239 and Feige~110 on 28 August 2011) and appeared stable within
1\%. The contamination starts around 6\,000\AA, as shown in
Figure~\ref{fig:NTTsecond}, and becomes severe around 6\,500\AA.

\subsection{TNG second order correction}

\begin{figure}
\centering
\includegraphics[angle=270,width=\columnwidth]{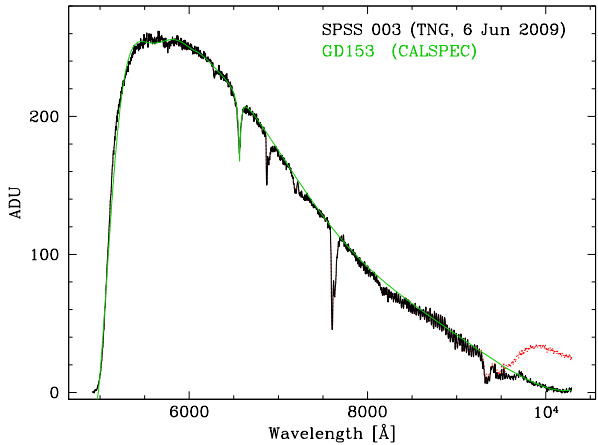} 
\caption{Same as Figure~\ref{fig:NTTsecond}, but for one TNG spectrum of GD~153,
taken in 2009 June. Apart from the telluric bands and the strong fringing that
was not corrected for at this stage, the second order contamination --- starting
around 9000-9500~\AA\  --- is very well corrected even with our simple
assumptions on the grism equation (see text).}
\label{fig:TNGsecond}
\end{figure}

The DOLORES red LR-R grism used in our SPSS campaign had a built-in
order-blocking filter for wavelengths bluer than $\simeq$5\,000~\AA, but
unfortunately some second-order contamination from blue light was observed after
$\simeq$9500~\AA\footnote{We have also observed an effect that could be ascribed
to second order contamination in the blue LR-B grism, starting at 6\,000~\AA\  
approximately. Because that region is always adequately covered by the red LR-R
grism, we simply cut the contaminated section of the LR-B spectra.} To obtain
the percentual contamination, $P$, we used two well known standards: HZ~44
\citep[a blue star,][]{landolt07} and G~146-76 \citep[a red star,][]{hog00}. A
third star, GD~153 \citep[another blue star,][]{bohlin95} was observed to
independently test the results. We also observed all LR-R spectra together with
the Johnson I broadband filter in front of the spectrograph, to recover
independently the uncontaminated spectral shape. They were observed with the
LR-B and LR-R grisms (see Table~\ref{tab:grisms}) both with a 5'' slit, the
widest available at the time, and with the 2'' slit for a better wavelength
calibration. We also tested GD~153 and KF06T2 \citep{reach05} on 2009 June 8
with similar results: the maximum difference between the two $P$ curves was
always lower than $\simeq$0.3\%. All spectra were extracted and wavelength
calibrated as in the NTT case, and reported to the same airmass with a tabulated
extinction
curve\footnote{http://www.ing.iac.es/astronomy/observing/manuals/ps/tech\_notes/\\tn031.pdf}.
We joined the spectra arbitrarily at 5850~\AA, i.e., well before the start of
second-order contamination. 

To determine a wavelength mapping relation, we obtained helium lamp spectra,
with 1~h t$_{\rm{exp}}$, with a B filter to block first order light. Of the
three measureable emission lines in LR-R, one was at 9553~\AA, where no He line
is expected. We thus could not derive a good fit of the function mapping the
second order, as done for the NTT case. We used the simple formula
$\lambda_1=2\lambda_2$, that derives from the simplest form of the grating
equation 
\begin{equation}
m\lambda=\sigma(\sin\beta\pm\sin\alpha)
\end{equation}
 We then derived the C$_1$
and C$_2$ curves and the percentage of contaminating flux using this basic
assumption: the emission line appearing at 9553~\AA\  corresponded well with the
4713~\AA\  He emission line. 

We tested the correction on TNG observations of various CALSPEC and literature
standards (one example is reported in Figure~\ref{fig:TNGsecond}). We obtained 
good results, having differences with SPSS literature spectra of 1--2\% at most,
for several stars. However, the presence of fringing and a deep telluric band
in the contaminated region (after 9500~\AA) causes high residuals, which can be
$\la$10\% in the worst cases. For many SPSS we had observations from other
telescopes, but for the few SPSS observed only with TNG, a viable solution will
be to replace the flux tables after 9500~\AA\  with model spectra \citep[see
also][]{bohlin07}.

\section{Conclusions}
\label{sec:concl}

To build a grid of $\sim$100 SPSS with $\leq$1\% internal errors and
$\leq$1--3\% external errors \citep[with respect to the Vega calibration
by][]{bohlin07} we have carried out a systematic study of instrumental effects
that could have an impact on the flux calibration of SPSS flux tables and
integrated magnitudes, finding methods to beat each effect separately down,
until their impact on fluxes (or ADUs) was below 1\%. We collected from the
literature the available methods and strategies to both evaluate and remove
those instrumental effects, and carried out specific daytime and nighttime
observations to quantify their effects. When necessary, we adapted the methods
to our case, or developed our specific tests and data-reduction methods.
Whenever feasible --- given the difficulties to obtain telescope time for such
programs --- we compared different approaches to the same problem. In
particular:

\begin{itemize}
\item{we characterized the employed CCDs to evaluate their minimum acceptable
exposure time, the shutter delay, and the linearity limit, using different
methods, and deriving recommendations for our SPSS observing campaigns (see
Tables~\ref{tab:shutter} and \ref{tab:lin});}
\item{we carried out a calibration frames monitoring plan, that provided
recommendations on the minimum frequency of each calibration type observations
(see Table~\ref{tab:plan}) for our observations;}
\item{as part of our calibration frames monitoring, we devised a specific
procedure to correct our spectra for fringing, after manipulation of
spectroscopic flat fields \citep[following the strategy by][]{malumuth03} and
using the IRAF {\em telluric} task;}
\item{we have also quantitatively studied the effect of gravity-induced lamp
flexures at DOLORES@TNG; we found shifts of about 3 pixels, with 0.01-0.06
errors, corresponding to almost 8~\AA\  in the wavelength space, confirming that
for this type of instruments day-time lamps are not sufficiently accurate. If
night-time lamps required too much observing time, we acquired high S/N day-time
lamps and shifted them onto the night-time ones to perform the wavelength
calibration;}
\item{we studied the effect of instrument-induced polarization with both
NTT and TNG, the two instruments that should have the largest polarization in
our SPSS campaigns. We observed stars with polarization levels up to 4\% at
different instrument rotation angles, and concluded that even in the case of
SPSS with those intrinsic polarization levels, the flux calibration remained
stable within 1\%;}
\item{we tested and adapted to our case the method described by
\citet{sanchez06} to correct our spectra for second order contamination; our
method can be applied in photometric and grey nights with residuals generally
within 1-2\% for NTT and TNG.}
\end{itemize}

In summary, a few of the examined instrumental effects turned out to be
negligible, while we devised specific methods to correct for the remaining ones,
reducing their impact on the flux calibration of SPSS magnitudes and spectra to
$\leq$1\%. The methods can be applied to a wide range of observational programs
on similar instruments. 

The actual reduction and analysis of the SPSS spectra will be presented in a
subsequent paper, where we will further reduce the residuals of instrumental
effect corrections by combining several independent observations for each SPSS.
We will also replace the low S/N blue and red borders (roughly
$\leq$3800--4000~\AA\  and $\geq$9500~\AA) with theoretical or semiempirical
templates.

\section*{Acknowledgments}

We warmly thank the technical staff of the San Pedro M\'artir, Calar Alto,
Loiano, La Silla NTT and REM, and Roque de Los Muchachos TNG observatories. We
acknowledge the support of INAF (Istituto Nazionale di Astrofisica) and ASI
(Agenzia Spaziale Italiana), under contracts I/037/08/0, I/058/10/0, and
2014-025-R.0, dedicated to the Gaia mission and to the Italian participation to
the Gaia DPAC. This work was supported by the MINECO (Spanish Ministry of
Economy) - FEDER through grants AYA2012-39551-C02-01 and ESP2013-48318-C2-1-R. 

This paper is based on data obtained with the following facilities:
BFOSC@Cassini in Loiano, Italy (in two runs in 2009 May and 2010 August);
EFOSC2@NTT in La Silla, Chile (under program ID 182.D-0287); DOLORES@TNG in La
Palma, Spain (from proposal TAC37 in the AOT16 semester, with observations in
2008 January, and from the TNG archive); CAFOS@2.2m in Calar Alto, Spain (under
program IDs H10-2.2-042, F07-2.2-033, F08-2.2-043, and H08-2.2-041); LaRuca@1.5m
in San Pedro M\'artir, Mexico (under program IDs \#27 and \#28, with
observations in 2008 August and 2010 July). 

This research has made use of the SIMBAD database, operated at CDS, Strasbourg,
France \citep{wenger00}. Figures were prepared with SuperMongo
(http://www.astro.princeton.edu/~rhl/sm/), except for Figure~\ref{fig:iris}
that was prepared with SAOImage DS9, developed by the Smithsonian Astrophysical
Observatory (http://ds9.si.edu/site/Home.html). This research has made use of
NASA's Astrophysics Data System Bibliographic Services
(http://adsabs.harvard.edu/index.html).

\newpage





\begin{thebibliography}{99}

\bibitem[\protect\citeauthoryear{Altavilla et al.}{2011}]{altavilla11} 
Altavilla, G., Pancino, E., Marinoni, S., Cocozza, G., Bellazzini, M., Bragaglia, A., Carrasco, J.~M., Federici, L.: 
2011, Gaia Technical Report No. GAIA-C5-TN-OABO-GA-004

\bibitem[\protect\citeauthoryear{Altavilla et al.}{2012}]{altavilla12} 
Altavilla, G., Botticella, M.~T., Cappellaro, E., Turatto M.:  
2012, Ap\&SS~341, 163 

\bibitem[\protect\citeauthoryear{Altavilla et al.}{2014}]{altavilla14} 
Altavilla, G., Ragaini, S., Pancino, E., Cocozza, G., Bellazzini, M., Galleti, S.,  Marinoni, S.:
2014, Gaia Technical Report No. GAIA-C5-TN-OABO-GA-005

\bibitem[\protect\citeauthoryear{Bailer-Jones  et al.}{2013}]{bailerjones13}
Bailer-Jones, C.~A.~L.,  Andrae, R., Arcay, B., et al.: 
2013, A\&A~559, 74 

\bibitem[\protect\citeauthoryear{Baldry et al.}{1999}]{baldry99} 
Baldry, I.~K., Viskum, M., Bedding, T.~R., Kjeldsen, H., Frandsen, S.: 
1999,  MNRAS~302, 381 

\bibitem[\protect\citeauthoryear{Breckinridge \& Oppenheimer}{2004}]{breckinridge04} 
Breckinridge, J.~B., Oppenheimer, B.~R.: 
2004, ApJ~600, 1091 

\bibitem[\protect\citeauthoryear{Bohlin, Colina \& Finley}{1995}]{bohlin95}
Bohlin, R.~C., Colina, L., Finley, D.~S.: 1995, AJ~110, 1316 

\bibitem[\protect\citeauthoryear{Bohlin, Dickinson \& Calzetti}{2001}]{bohlin01} 
Bohlin, R.~C., Dickinson, M.~E., Calzetti, D.: 2001, AJ~122, 2118 

\bibitem[\protect\citeauthoryear{Bohlin}{2007}]{bohlin07} 
Bohlin,  R.~C.: 
2007, ASPC~364, 315 

\bibitem[\protect\citeauthoryear{Bohlin, Gordon, \& Tremblay}{2014}]{bohlin14}
Bohlin, R.~C., Gordon, K.~D., Tremblay, P.-E.:
2014, PASP~126, 711 

\bibitem[\protect\citeauthoryear{Busonero et al.}{2006}]{busonero06} 
Busonero, D., Gai, M., Gardiol, D., Lattanzi, M.~G., Loreggia, D.:
2006, A\&A~449, 827 

\bibitem[\protect\citeauthoryear{Carrasco et al.}{2007}]{carrasco07} 
Carrasco, J.~M., Jordi, C., Lopez-Marti, B., Figueras, F., Anglada-Escude, G.:
2007, Gaia Technical Report No. GAIA-C5-TN-UB-JMC-002

\bibitem[\protect\citeauthoryear{Cocozza et al.}{2013}]{cocozza13} 
Cocozza, G., Altavilla, G., Carrasco, J.~M., Pancino, E., Marinoni, S., Galleti, S.:
2013, Gaia Technical Report No. GAIA-C5-TN-OABO-GCC-001 

\bibitem[\protect\citeauthoryear{Djorgovski \& Dickinson}{1989}]{djorgovski89}
Djorgovski, S., Dickinson, M.:
1989, HiA~8, 645 

\bibitem[\protect\citeauthoryear{Gilliland et al.}{1993}]{gilliland93} 
Gilliland R.~L., Brown, T.~M., Kjeldsen, H., et al.:
1993, AJ~106,  2441 

\bibitem[\protect\citeauthoryear{Giro et al.}{2003}]{giro03} 
Giro, E., Bonoli, C., Leone, F., Molinari, E., Pernechele, C., Zacchei, A.:
2003, SPIE~4843, 456 

\bibitem[\protect\citeauthoryear{Gullixson}{1992}]{gullixson92} 
Gullixson, C.~A.:
1992, ASPC~23, 130 

\bibitem[\protect\citeauthoryear{Gutierrez-Moreno et al.}{1994}]{gutierrez94}
Gutierrez-Moreno, A., Heathcote, S., Moreno,  H., Hamuy M.:
1994, PASP~106, 1184 

\bibitem[\protect\citeauthoryear{Jordi et al.}{2010}]{jordi10} 
Jordi, C.,   Gebran, M.,  Carrasco, J. M.,  et al.:
2010, A\&A~523, 48 

\bibitem[\protect\citeauthoryear{Hamuy et al.}{1992}]{hamuy92}  
Hamuy, M., Walker, A.~R., Suntzeff, N.~B., Gigoux, P., Heathcote, S.~R.,  Phillips, M.~M.: 
1992, PASP~104, 533 

\bibitem[\protect\citeauthoryear{Hamuy et al.}{1994}]{hamuy94}  
Hamuy, M., Suntzeff, N.~B., Heathcote, S.~R., Walker, A.~R., Gigoux, P.,  Phillips, M.~M.:
1994, PASP~106, 566 

\bibitem[\protect\citeauthoryear{Howell}{2012}]{howell12} 
Howell, S.~B.:
2012, PASP~124, 263 

\bibitem[\protect\citeauthoryear{H{\o}g et al.}{2000}]{hog00} 
H{\o}g, E., Fabricius, C., Makarov, V. V., et al.:
2000, A\&A~357, 367 

\bibitem[\protect\citeauthoryear{Koch et al.}{2004}]{koch04} 
Koch, A., Odenkirchen, M., Grebel, E.~K., Caldwell, J.~A.~R.:
2004, AN~325, 299 

\bibitem[\protect\citeauthoryear{Landolt}{1992}]{landolt92} 
Landolt, A.~U.:
1992, AJ~104, 340 

\bibitem[\protect\citeauthoryear{Landolt  \& Uomoto}{2007}]{landolt07} 
Landolt, A.~U., Uomoto, A.~K.:
2007, AJ~133, 768 

\bibitem[\protect\citeauthoryear{Leach et al.}{1980}]{leach80} 
Leach, R.~W., Schild, R.~E., Gursky, H., Madejski, G.~M., Schwartz, D.~A., Weekes, T.~C.: 
1980, PASP~92, 233 

\bibitem[\protect\citeauthoryear{Leach}{1987}]{leach87} 
Leach, R.~W.:
1987, OptEn~26, 1061 

\bibitem[\protect\citeauthoryear{Lesser}{1990}]{lesser90} 
Lesser, M.~P.:
1990, ASPC~8, 65 

\bibitem[\protect\citeauthoryear{Lindegren et al.}{2008}]{lindegren08} 
Lindegren, L., Babusiaux, C., Bailer-Jones, C., et al.:
2008, IAUS~248, 217 

\bibitem[\protect\citeauthoryear{Malumuth et al.}{2003}]{malumuth03} 
Malumuth, E.~M., Hill, R.~S., Gull, T.,  et al.: 
2003, PASP~115,  218 

\bibitem[\protect\citeauthoryear{Manfroid, Selman, \& Jones}{2001}]{manfroid01}
Manfroid, J., Selman, F., Jones, H.:
2001, Msngr~104, 16 


\bibitem[\protect\citeauthoryear{Marinoni}{2011}]{marinoni11} 
Marinoni, S.:
2011, PhDT, Bologna University 

\bibitem[\protect\citeauthoryear{Marinoni et al.}{2012}]{marinoni12} 
Marinoni, S., Pancino, E., Altavilla, G., Cocozza, G., Carrasco, J.~M., Mongui\'o, M.,  Vilardell, F.: 
2012, Gaia Technical Report No. GAIA-C5-TN-OABO-SMR-001

\bibitem[\protect\citeauthoryear{Marinoni et al.}{2013}]{marinoni13} 
Marinoni, S., Galleti, S., Cocozza, G., Pancino, E., Altavilla, G.: 
2013, Gaia Technical Report No. GAIA-C5-TN-OABO-SMR-002

\bibitem[\protect\citeauthoryear{Massey}{1997}]{massey97} 
Massey, P.:
1997, iraf.noao.edu/iraf/docs/ccduser3.ps.Z

\bibitem[\protect\citeauthoryear{Massey \& Hanson}{2013}]{massey13} 
Massey, P., Hanson, M.~M.:
2013, pss2.book, 35 

\bibitem[\protect\citeauthoryear{Mignard}{2005}]{mignard05} 
Mignard, F.: 
2005, ASPC~338, 15 

\bibitem[\protect\citeauthoryear{Mignard et al.}{2008}]{mignard08} 
Mignard, F., Bailer-Jones, C., Bastian, U., et al.: 
2008, IAUS~248, 224 

\bibitem[\protect\citeauthoryear{Munari \& Lattanzi}{1992}]{munari92} 
Munari, U., Lattanzi, M.~G.:
 1992, PASP~104, 121 

\bibitem[\protect\citeauthoryear{Newberry}{1991}]{newberry91} 
Newberry, M.~V.:
1991, PASP~103, 122 

\bibitem[\protect\citeauthoryear{Oke}{1990}]{oke90} 
Oke, J.~B.:
1990, AJ~99, 1621 

\bibitem[\protect\citeauthoryear{Pak{\v s}tiene \& Solheim}{2003}]{pakstiene03}
Pak{\v s}tiene, E., Solheim, J.-E.: 
2003, BaltA~12, 221 

\bibitem[\protect\citeauthoryear{Pancino et al.}{2008}]{pancino08} 
Pancino, E., Altavilla, G., Bellazzini, M., Marinoni, S., Bragaglia, A., Federici, L., Cacciari, C.:
2008, Gaia Technical Report No. GAIA-C5-TN-OABO-EP-001

\bibitem[\protect\citeauthoryear{Pancino et al.}{2009}]{pancino09} 
Pancino, E., Altavilla, G., Carrasco, J.~M., et al.: 
2009, Gaia Technical Report No. GAIA-C5-TN-OABO-EP-003

\bibitem[\protect\citeauthoryear{Pancino et al.}{2011}]{pancino11} 
Pancino, E., Altavilla, G., Carrasco, J.~M., Marinoni, S., Cocozza, G., Bellazzini, M., Federici, L.:
2011, Gaia Technical Report No. GAIA-C5-TN-OABO-EP-006

\bibitem[\protect\citeauthoryear{Pancino et al.}{2012}]{pancino12} 
Pancino E., Altavilla, G., Marinoni, S., et al.: 
2012, MNRAS~426, 1767 

\bibitem[\protect\citeauthoryear{Perryman et al.}{2001}]{perryman01} 
Perryman, M.~A.~C., de Boer, K.~S., Gilmore, G., et al.: 
2001, A\&A~369, 339 

\bibitem[\protect\citeauthoryear{Prusti}{2011}]{prusti11} 
Prusti T.:
2011, EAS~45, 9 

\bibitem[\protect\citeauthoryear{Prusti}{2012}]{prusti12} 
Prusti, T.: 
2012, AN~333, 453 

\bibitem[\protect\citeauthoryear{Reach et al.}{2005}]{reach05} 
Reach W.~T., Megeath, S. T., Cohen, M., et al.: 
2005, PASP~117, 978 

\bibitem[\protect\citeauthoryear{S{\'a}nchez-Bl{\'a}zquez et al.}{2006}]{sanchez06} 
S{\'a}nchez-Bl{\'a}zquez P., Peletier, R.~F., Jim{\'e}nez-Vicente, J.,  et al.: 
2006, MNRAS~371,703 

\bibitem[\protect\citeauthoryear{Wenger et al.}{2000}]{wenger00} 
Wenger M., Ochsenbein, F., Egret, D., et al.: 
2000, A\&AS~143, 9 

\bibitem[\protect\citeauthoryear{Sch{\"o}nebeck et al.}{2014}]{schonebeck14}
Sch{\"o}nebeck, F., Puzia, T.~H., Pasquali, A., Grebel, E.~K., Kissler-Patig, M., Kuntschner, H., Lyubenova, M., Perina, S.: 
2014, A\&A~572, 13 

\bibitem[\protect\citeauthoryear{Soubiran et al.}{2013}]{soubiran13} 
Soubiran, C., Jasniewicz, G., Chemin, L., Crifo, F., Udry, S., Hestroffer, D., Katz, D.:
2013, A\&A~552, 64 

\bibitem[\protect\citeauthoryear{Stello et al.}{2006}]{stello06} 
Stello D., Arentoft, T., Bedding, T. R, et al.: 
2006, MNRAS~373, 1141 

\bibitem[\protect\citeauthoryear{Sullivan et al.}{2011}]{sullivan11} 
Sullivan, M., Guy, J., Conley, A., et al.: 
2011, ApJ~737, 102 

\bibitem[\protect\citeauthoryear{Szokoly et al.}{2004}]{szokoly04} 
Szokoly, G.~P., Bergeron, J., Hasinger, G.,  et al.: 
2004, ApJS~155,  271 

\bibitem[\protect\citeauthoryear{Traub}{1990}]{traub90} 
Traub, W.~A.: 
1990, JOSAA~7, 1779 

\bibitem[\protect\citeauthoryear{Turnshek et al.}{1990}]{turnshek90} 
Turnshek, D.~A., Bohlin, R.~C., Williamson,  R.~L.~II, Lupie, O.~L., Koornneef, J., Morgan, D.~H.:
1990, AJ~99, 1243 

\bibitem[\protect\citeauthoryear{Tyson}{1990}]{tyson90} 
Tyson J.~A.: 
1990, ASPC~8, 1 

\bibitem[\protect\citeauthoryear{van Leeuwen \& Richards}{2012}]{FVL-001} 
van Leeuwen, F., Richards, P.: 
2012, Gaia Technical Report No. GAIA-C5-TN-IOA-FVL-001

\bibitem[\protect\citeauthoryear{Walker}{1993}]{walker93} 
Walker, A.~R.: 
1993, spct.conf, 278 



\end{thebibliography}
\end{document}